\documentclass[12pt]{article} 
 
\usepackage{amsmath,amssymb}
\usepackage{amsthm}        
\usepackage{curves}

\theoremstyle{definition}\newtheorem{definition}{Definition}
\theoremstyle{plain}\newtheorem{lemma}{Lemma} 
\theoremstyle{plain}\newtheorem{proposition}{Proposition}

 \font\tenscr=rsfs10 scaled1100
 \font\sevenscr=rsfs7 % scaled \magstep1
 \font\fivescr=rsfs5 % scaled \magstep1
 \skewchar\tenscr='177 \skewchar\sevenscr='177 \skewchar\fivescr='177
 \newfam\scrfam
 \textfont\scrfam=\tenscr
 \scriptfont\scrfam=\sevenscr
 \scriptscriptfont\scrfam=\fivescr

 \def\J{{\fam\scrfam I}}

\newcommand{\M}{{\mathcal{M}}}
\newcommand{\s}{{\mathcal{S}}}
\newcommand{\T}{{\mathcal{T}}}
\newcommand{\D}{{\mathcal{D}}}
\newcommand{\C}{{\mathcal{C}}}
\newcommand{\K}{{\mathcal{K}}}
\newcommand{\N}{{\mathcal{N}}}
\newcommand{\U}{{\mathcal{U}}}
\newcommand{\V}{{\mathcal{V}}}
\newcommand{\W}{{\mathcal{W}}}
\newcommand{\B}{{\mathcal{B}}}
\newcommand{\R}{{\mathbb{R}}}
\newcommand{\n}{{\mathbb{N}}}
\newcommand{\F}{{\mathbb{F}}}
\newcommand{\g}{{\mathrm{deg}}}

\begin{document}

%\begin{flushright}
%	submitted to COMP
%    	September 2000
%\end{flushright}
$\,$
\vspace{1cm}

\begin{center}
{\Large{\bf Global properties of gravitational lens maps in a 
\\[0.1cm]
Lorentzian manifold setting}}

\vspace{0.4cm}

{\large{\em Volker Perlick}}
\vspace{0.3cm}

{\large{Albert Einstein Institute, 14476 Golm, Germany}} \\
\vspace{0.15cm}

Permanent address: TU Berlin, Sekr. PN 7-1, 10623 Berlin, Germany \\
{\it email\/}: vper0433@itp.physik.tu-berlin.de
\end{center}

%%%%%%%%%%%%%%%%%%%%%%%%%%%
\begin{abstract}
	In a general-relativistic spacetime (Lorentzian manifold),
	gravitational lensing can be characterized by a lens map,
	in analogy to the lens map of the quasi-Newtonian 
	approximation formalism. The lens map is defined on 
	the celestial sphere of the observer (or on part of it)
	and it takes values in a two-dimensional manifold 
	representing a two-parameter family of worldlines. In this
	article we use methods from differential topology to 
	characterize global properties of the lens map. Among 
	other things, we use the mapping degree (also known as
	Brouwer degree) of the lens map as a tool for characterizing
	the number of images in gravitational lensing situations.
	Finally, we illustrate the general results with gravitational
	lensing (a) by a static string, (b) by a spherically
	symmetric body, (c) in asymptotically simple and empty 
	spacetimes, and (d) in weakly perturbed Robertson-Walker 
	spacetimes.
\end{abstract}

%%%%%%%%%%%%%%%%%%%%%%%%%%%%%%%%%%%%%%%%%%%%%%%%%%%%%%%%%%%%%%%%%%%%%%%%%

\section{Introduction}\label{sec:intro}
	Gravitational lensing is usually studied in a quasi-Newtonian 
	approximation formalism which is essentially based on the assumptions
	that the gravitational fields are weak and that the bending angles
	are small, see Schneider, Ehlers and Falco \cite{SEF} for a 
	comprehensive discussion. This formalism has proven to be very 
	powerful for the calculation of special models. In addition it has 
	also been used for proving general theorems on the qualitative 
	features of gravitational lensing such as the possible number of 
	images in a multiple imaging situation. As to the latter point, it 
	is interesting to inquire whether the results can be reformulated 
	in a Lorentzian manifold setting, i.e., to inquire to what extent 
	the results depend on the approximations involved. 

	In the quasi-Newtonian approximation formalism one considers light 
	rays in Euclidean 3-space that go from a fixed point (observer) to a 
	point that is allowed to vary over a 2-dimensional plane (source
	plane). The rays are assumed to be straight lines with the only
	exception that they may have a sharp bend at a 2-dimensional plane
	(deflector plane) that is parallel to the source plane. (There is
	also a variant with several deflector planes to model deflectors 
	which are not ``thin''.) For each concrete mass distribution, the 
	deflecting angles are to be calculated with the help of Einstein's
	field equation, or rather of those remnants of Einstein's field
	equation that survive the approximations involved. Hence, at 
	each point of the deflector plane the deflection angle is uniquely
	determined by the mass distribution. As a consequence,
	following light rays from the observer into the past always gives 
	a unique ``lens map'' from the deflector plane to the source plane. 
	There is ``multiple imaging'' whenever this lens map fails to
	be injective.

	In this article we want to inquire whether an analogous lens map 
	can be introduced in a spacetime setting, without using 
	quasi-Newtonian approximations. According to the rules of
	general relativity, a spacetime is to be modeled by a 
	Lorentzian manifold $(\M,g)$ and the light rays are to be 
	modeled by the lightlike geodesics in $\M$. We shall assume that 
	$(\M,g)$ is time-oriented, i.e., that the timelike and lightlike
	vectors can be distinguished into future-pointing and past-pointing
	in a globally consistent way. To define a general 
	lens map, we have to fix a point $p \in \M$ as the event where 
	the observation takes place and we have to look for 
	an analogue of the deflector plane and for an analogue of the 
	source plane. As to the deflector plane, there is an obvious
	candidate, namely the {\em celestial sphere\/} $\s_p$ at $p$. 
	This can be defined as the the set of all one-dimensional 
	lightlike subspaces of the tangent space $T_p\M$ or, equivalently,
	as the totality of all light rays issuing from $p$ into the past. 
	As to the source plane, however, there is no 
	natural candidate. Following Frittelli, Newman and Ehlers 
	\cite{FN,EFN, Eh}, one might consider any timelike 3-dimensional 
	submanifold $\T$ of the spacetime manifold as a substitute for 
	the source plane. The idea is to view such a submanifold as 
	ruled by worldlines of light sources. To make this 
	more explicit, one could restrict to the case that 
	$\T$ is a fiber bundle over a two-dimensional manifold 
	$\N$, with fibers timelike and diffeomorphic to $\R$. 
	Each fiber is to be interpreted as the worldline
	of a light source, and the set $\N$ may be identified with
	the set of all those worldlines. In this situation
	we wish to define a lens map $f_p : \s_p \longrightarrow \N$ 
	by extending each light ray from $p$ into the past
	until it meets $\T$ and then projecting onto $\N$. In general,
	this prescription does not give a well-defined map since 
	neither existence nor uniqueness of the target value is guaranteed.
	As to existence, there might be some past-pointing lightlike 
	geodesics from $p$ that never reach $\T$. As to uniqueness, 
	one and the same light ray might intersect $\T$ several times. 
	The uniqueness problem could be circumvented by considering, 
	on each past-pointing lightlike geodesic from $p$, only the 
	first intersection with $\T$, thereby willfully excluding some 
	light rays from the discussion. This comes up to ignoring every 
	image that is hidden behind some other image of a light source 
	with a worldline $\xi \in \N$. For the existence problem, however, 
	there is no general solution. Unless one restricts to special 
	situations, the lens map will be defined only on some subset 
	$\D_p$ of $\s_p$ (which may even be empty). Also, one would 
	like the lens map to be differentiable or at least continuous. 
	This is guaranteed if one further restricts the domain $\D_p$ 
	of the lens map by considering only light rays that meet 
	$\T$ transversely. 

	Following this line of thought, we give a precise definition
	of lens maps in Section~\ref{sec:deff}. We will be a little bit
	more general than outlined above insofar as the source surface
	need not be timelike; we also allow for the limiting case of a 
	lightlike source surface. This has the advantage that we may
	choose the source surface ``at infinity'' in the 
	case of an asymptotically simple and empty spacetime. In
	Section~\ref{sec:regular} we briefly discuss some general
	properties of the caustic of the lens map. In 
	Section~\ref{sec:degree} we introduce the mapping degree (Brouwer 
	degree) of the lens map as an important tool from differential
	topology. This will then give us some theorems on the 
	possible number of images in gravitational lensing 
	situations, in particular in the case that we have a ``simple 
	lensing neighborhood''. The latter notion will be introduced and 
	discussed in Section~\ref{sec:sln}. We conclude with 
	applying the general results to some examples in 
	Section~\ref{sec:examples}.

	Our investigation will be purely geometrical in
	the sense that we discuss the influence of the spacetime 
	geometry on the propagation of light rays but not the influence
	of the matter distribution on the spacetime geometry. In other
	words, we use only the geometrical background of general
	relativity but not Einstein's field equation. For this reason
	the ``deflector'', i.e., the matter distribution that is the
	cause of gravitational lensing, never explicitly appears in
	our investigation. However, information on whether the deflectors
	are transparent or non-transparent will implicitly enter into
	our considerations.

%%%%%%%%%%%%%%%%%%%%%%%%%%%%%%%%%%%%%%%%%%%%%%%%%%%%%%%%%%%%%%%%%%%%%%%%%%%

\section{Definition of the lens map}\label{sec:deff}
	As a preparation for precisely introducing the lens map
	in a spacetime setting, we first specify some terminology.

	By a {\em manifold\/} we shall always mean what is more fully
	called a ``real, finite-dimensional, Hausdorff, second
	countable (and thus paracompact) $C^{\infty}$-manifold without 
	boundary''. Whenever we have a $C^{\infty}$ vector field $X$ on 
	a manifold $\M$, we may consider two points in $\M$ as 
	{\em equivalent\/} if they lie on the same integral curve of 
	$X$. We shall denote the resultant quotient space, which may 
	be identified with the set of all integral curves of $X$, by 
	$\M /X$. We call $X$ a {\em regular\/} vector field if $\M /X$ 
	can be given the structure of a manifold in such a way that 
	the natural projection $\pi_X : \M \longrightarrow \M /X$
	becomes a $C^{\infty}$-submersion. It is easy to construct 
	examples of non-regular vector fields. E.g., if $X$ has no zeros
	and is defined on $\R ^n \setminus \{0\}$, then $\M /X$ cannot
	satisfy the Hausdorff property, so it cannot be a manifold 
	according to our terminology. Palais \cite{Pa} has proven a
	useful result which, in our terminology, can be phrased in the
	following way. If none of $X$'s integral curves
	is closed or almost closed, and if $\M /X$ satisfies the 
	Hausdorff property, then $X$ is regular.

	We are going to use the following terminology. A {\em 
	Lorentzian manifold\/} is a manifold $\M$ together with a
	$C^{\infty}$ metric tensor field $g$ of Lorentzian signature 
	$(+ \dots  + -)$. A Lorentzian manifold is {\em time-orientable\/}
	if the set of all timelike vectors $\{ Z \in T \M \, | \, 
	g(Z,Z) < 0\}$ has exactly two connected components. Choosing
	one of those connected components as {\em future-pointing\/} 
	defines a {\em time-orientation\/} for 
	$(\M, g)$. A {\em spacetime\/} is a connected 4-dimensional 
	time-orientable Lorentzian manifold together with a 
	time-orientation.

	We are now ready to define what we will call a ``source surface'' 
	in a spacetime. This will provide us with the target space for 
	lens maps.
\begin{definition}\label{def:TW}
	$(\T,W)$ is called a {\em source surface\/} in a spacetime
	$(\M,g)$ if 
\\
	(a) $\T$ is a 3-dimensional $C^{\infty}$ submanifold of $\M$;
\\
	(b) $W$ is a nowhere vanishing regular $C^{\infty}$ vector 
	field on $\T$ which is everywhere causal, $g(W,W) \le 0$,
	and future-pointing;
\\
	(c) $\pi _W : \T \longrightarrow \N = \T /W$ is a fiber bundle
	with fiber diffeomorphic to $\R$ and the quotient manifold 
	$\N = \T /W$ is connected and orientable.
\end{definition}
	We want to interpret the integral curves of $W$ as the worldlines
	of light sources. Thus, one should assume that they are not
	only causal but even timelike, $g(W,W) < 0$, since a light 
	source should move at subluminal velocity. For technical
	reasons, however, we allow for the possibility that an integral
	curve of $W$ is lightlike (everywhere or at some points), 
	because such curves may appear as ($C^1$-)limits of 
	timelike curves. This will give us the possibility to apply 
	the resulting formalism to asymptotically simple and empty 
	spacetimes in a convenient way, see 
	Subsection~\ref{subsec:asy} below. Actually, the causal character 
	of $W$ will have little influence upon the results we want 
	to establish. What really matters is a transversality 
	condition that enters into the definition of the lens map below. 

	Please note that, in the situation of 
	Definition~\ref{def:TW}, the bundle $\pi _W : \T \longrightarrow 
	\N$ is necessarily trivializable, i.e., $\T \simeq \N \times 
	\R$. To prove this, let us assume that the flow of $W$ is defined 
	on all of $\R \times \T$, so it makes $\pi _W : \T \longrightarrow 
	\N$ into a principal fiber bundle. (This is no restriction of 
	generality since it can always be achieved by multiplying $W$ with
	an appropriate function. This function can be determined in the 
	following way. Owing to a famous theorem of Whitney \cite{Whi},
	also see Hirsch \cite{Hi}, p.55, paracompactness guarantees 
	that $\T$ can be embedded as a closed submanifold into $\R ^n$ 
	for some $n$. Pulling back the Euclidean metric gives a complete
	Riemannian metric $h$ on $\T$ and the flow of the vector field
	$h(W,W)^{-1/2}W$ is defined on all of $\R \times \T$, cf. Abraham and 
	Marsden \cite{AM}, Proposition~2.1.21.) 
	Then the result follows from the well known facts that any fiber 
	bundle whose typical fiber is diffeomorphic to $\R ^n$ admits a 
	global section (see, e.g., Kobayashi and Nomizu \cite{KN}, p.58), 
	and that a principal fiber bundle is trivializable if and only 
	if it admits a global section (see again \cite{KN}, p.57).

	Also, it is interesting to note the following. If $\T$ is any 
	3-dimensional submanifold of $\M$ that is foliated into timelike
	curves, then time orientability guarantees that these are the 
	integral curves of a timelike vector field $W$. If we assume, 
	in addition, that $\T$ contains no closed timelike curves, then 
	it can be shown that $\pi _W : \T \longrightarrow \N$ is 
	necessarily a fiber bundle with fiber diffeomorphic to $\R$, 
	providing $\N$ satisfies the Hausdorff property, see Harris 				\cite{Ha}, Theorem~2. This shows that there is 
	little room for relaxing the conditions of 
	Definition~\ref{def:TW}.

	Choosing a source surface in a spacetime will give us the 
	target space $\N = \T /W$ for the lens map. To specify the 
	domain of the lens map, we consider, at any point $p \in \M$, the set 
	$\s_p$ of all lightlike directions at $p$, i.e., the set of
	all one-dimensional lightlike subspaces of $T_p \M$. We shall 
	refer to $\s_p$ as to the {\em celestial sphere\/} at $p$. This 
	is justified since, obviously, $\s_p$ is in natural one-to-one 
	relation with the set of all light rays arriving at $p$. As it is 
	more convenient to work with vectors rather than with directions, 
	we shall usually represent $\s_p$ as a submanifold
	of $T_p \M$. To that end we fix a future-pointing timelike 
	vector $V_p$ in the tangent space $T_p\M$. The vector $V_p$ may be 
	interpreted as the 4-velocity of an observer at $p$. We 
	now consider the set
\begin{equation}\label{eq:Sp}
	\s _p = \big\{ Y_p \in T_p \M \, \big| \, g(Y_p,Y_p)=0 \;
	{\mathrm{and}} \; g(Y_p,V_p) = 1 \; \big\} \, .
\end{equation}
	It is an elementary fact that (\ref{eq:Sp}) defines an embedded
	submanifold of $T_p \M$ which is diffeomorphic to the standard 
	2-sphere $S^2$. As indicated by our notation, the set (\ref{eq:Sp}) 
	can be identified with the celestial sphere at $p$, just by relating
	each vector to the direction spanned by it. 

	Representation (\ref{eq:Sp}) of the celestial sphere gives a 
	convenient way of representing the light rays through $p$. We 
	only have to assign to each $Y_p \in \s_p$ the lightlike geodesic
	$s \longmapsto {\mathrm{exp}}_p ( s Y_p)\,$ 
	where exp$_p : W_p \subseteq T_p \M \longrightarrow \M$ denotes 
	the exponential map at the point $p$ of the Levi-Civita connection 
	of the metric $g$. Please note that this geodesic is past-pointing, 
	because $V_p$ was chosen future-pointing, and that it passes 
	through $p$ at the parameter value $s=0$. 

\begin{figure}
\begin{center}
\setlength{\unitlength}{1cm}
\begin{picture}(10,10)
	\curve(6,0.5, 5,1, 4,1.2, 3,1.2)
  	\curve(6,3.5, 5,4, 4,4.2, 3,4.2)
	\curve(6,8.5, 5,9, 4,9.2, 3,9.2)
	\curve(6,3.5, 6,8.5)
	\curve(5,4, 5,9)
	\curve(4,4.2, 4,9.2)
	\curve(3,4.2, 3,9.2)
        \curve(1.5,8, 4.5,5.2)        
	\put(4.7,7.5){\vector(0,1){1}}
	\put(1.5,8){\circle*{0.08}}
	\put(1.5,8){\vector(1,-1){0.5}}
	\put(4.5,5.2){\circle*{0.08}}
 	\put(4.5,1.13){\circle*{0.08}}     
	\put(6.4,7){${\mathcal{T}}$}
	\put(6.4,0.5){${\mathcal{N}}$}
	\put(1.2,8.3){$p$}
        \put(1.4,7.2){$Y_p$}
	\put(4.1,8.){$W$}
	\put(3.9,0.6){$f_p(Y_p)$}
	\put(4.5,3.2){\vector(0,-1){1}}
	\put(4.6,2.6){$\pi _W$}
\end{picture}
\end{center}
\caption{Illustration of the lens map}\label{fig:fp} 
\end{figure}
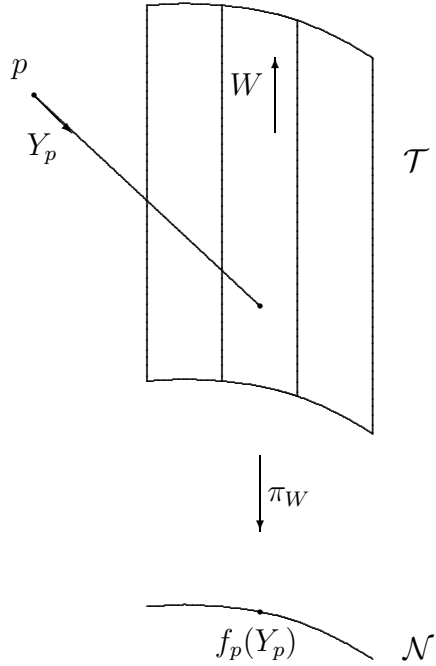

	The lens map is defined in the following way. After fixing a
	source surface $(\T,W)$ and choosing a point $p \in \M$, we
	denote by $\D_p \subseteq \s_p$ the subset of all lightlike
	directions at $p$ such that the geodesic to which this direction
	is tangent meets $\T$ (at least once) if sufficiently extended
	to the past, and if at the first intersection point $q$ with 
	$\T$ this geodesic is transverse to $\T$. By projecting $q$ to 
	$\N = \T /W$ we get the lens map $f_p : \D_p \longrightarrow 
	\N = \T /W$, see Figure~\ref{fig:fp}. If we use the 
	representation (\ref{eq:Sp}) for $\s_p$, the definition of the
	lens map can be given in more formal terms in the following way.
\begin{definition}\label{def:lensmap}
	Let $(\T,W)$ be a source surface in a spacetime $(\M,g)$. Then, 
	for each $p \in \M$, the lens map $f_p : \D_p \longrightarrow 
	\N = \T /W$ is defined in the following way. In the notation
	of equation (\ref{eq:Sp}), let $\D_p$ be the set of all 
	$Y_p \in \s_p$ such that there is a real number $w_p(Y_p) > 0$ 
	with the properties
\\
	(a) $sY_p$ is in the maximal domain of the exponential map
	for all $s \in [\, 0 \, ,w_p(Y_p)]$;
\\
	(b) the curve $s \longmapsto {\mathrm{exp}} (sY_p)$ intersects 
	$\T$ at the value $s = w_p(Y_p)$ transversely;
\\
	(c) ${\mathrm{exp}}_p ( s Y_p) \notin \T$ for all 
	$s \in [\, 0 \, , w_p(Y_p)[\;$.
\\
	This defines a map $w_p : \D_p \longrightarrow \R$. The {\em lens 
	map\/} at $p$ is then, by definition, the map
\begin{equation}\label{eq:lensmap}
	f_p : \D_p \longrightarrow \N = \T/X \, , \quad
	f_p(Y_p) = \pi_W \big( {\mathrm{exp}}_p ( w_p(Y_p )Y_p)\big) \, .
\end{equation}
	Here $\pi_W : \T \longrightarrow \N$ denotes the natural 
	projection.
\end{definition}

	The transversality condition in part (b) of 
	Definition~\ref{def:lensmap} guarantees that the domain 
	$\D_p$ of the lens map is an open subset of $\s_p$. The 
	case $\D_p = \emptyset$ is, of course, not excluded. In 
	particular, $\D_p = \emptyset$ whenever $p \in \T$, owing 
	to part (c) of Definition~\ref{def:lensmap}.

	Moreover, the transversality condition in part (b) of 
	Definition~\ref{def:lensmap}, in combination with the
	implicit function theorem, makes sure that the map $w_p :
	\D_p \longrightarrow \R$ is a $C^{\infty}$ map. As the 
	exponential map of a $C^{\infty}$ metric is again $C^{\infty}$,
	and $\pi_W$ is a $C^{\infty}$ submersion by assumption, this
	proves the following.  
\begin{proposition}\label{prop:diff}
	The lens map is a $C^{\infty}$ map.
\end{proposition}

	Please note that without the transversality condition the lens
	map need not even be continuous. 

	Although our Definition~\ref{def:lensmap} made use of the 
	representation (\ref{eq:Sp}), which refers to a timelike 
	vector $V_p$, the lens map is, of course, independent of 
	which future-pointing $V_p$ has been chosen. We decided 
	to index the lens map only with $p$ although, strictly 
	speaking, it depends on $\T$, on $W$, and on $p$. Our 
	philosophy is to keep a source surface $(\T,W)$ fixed, 
	and then to consider the lens map for all points $p \in \M$. 

	In view of gravitational lensing, the lens map admits the 
	following interpretation. 
	For $\xi \in \N$, each point $Y_p \in \D_p$ with $f_p (Y_p)$
	corresponds to a past-pointing lightlike geodesic from $p$
	to the worldline $\xi$ in $\M$, i.e., it corresponds to an
	image at the celestial sphere of $p$ of the light source with
	worldline $\xi$. If $f_p$ is not injective, we are in a multiple
	imaging situation. The converse need not be true as the lens
	map does not necessarily cover all images. There might be 
	a past-pointing lightlike geodesic from $p$ reaching $\xi$ 
	after having met $\T$ before, or being tangential to $\T$ on
	its arrival at $\xi$. In either case, the corresponding image
	is ignored by the lens map. The reader might be inclined to
	view this as a disadvantage. However, in 
	Section~\ref{sec:examples} below we  discuss some situations 
	where the existence of such additional light rays can be 
	excluded (e.g., asymptotically simple and empty spacetimes)
	and situations where it is desirable, on physical grounds, to
	disregard such additional light rays (e.g., weakly perturbed 
	Robertson-Walker spacetimes with compact spatial sections).

	It was already mentioned that the domain $\D_p$ of the lens 
	map might be empty; this is, of course, the worst case that
	could happen. The best case is that the domain is all of
	the celestial sphere, $\D_p = \s_p$. We shall see in the 
	following sections that many interesting results are true 
	just in this case. However, there are several cases of 
	interest where $\D_p$ is a proper subset of $\s_p$.
	If the domain of the lens map $f_p$ is the whole celestial 
	sphere, none of the light rays issuing from $p$ into the 
	past is blocked or trapped before it reaches $\T$. In view
	of applications to gravitational lensing, this excludes 
	the possibility that these light rays meet a non-transparent
	deflector. In other words, it is a typical feature of 
	gravitational lensing situations with non-transparent
	deflectors that $\D_p$ is not all of $\s_p$. Two simple 
	examples, viz., a non-transparent string and a non-transparent 
	spherical body, will be considered in 
	Subsection~\ref{subsec:string} below.

%%%%%%%%%%%%%%%%%%%%%%%%%%%%%%%%%%%%%%%%%%%%%%%%%%%%%%%%%%%%%%%%%%%%%%%

\section{Regular and critical values of the lens map}\label{sec:regular}
	Please recall that, for a differentiable map $F : \M_1 
	\longrightarrow \M_2$ between two manifolds, $Y \in \M_1$ 
	is called a {\em regular point\/} of $F$ if the differential
	$T_{Y} F : T_{Y} \M_1 \longrightarrow T_{F(Y)} \M_2$ 
	has maximal rank, otherwise $Y$ is called a {\em critical point}.
	Moreover, $\xi \in \M_2$ is called a {\em regular value\/} of 
	$F$ if all $Y \in F^{-1} (\xi )$ are regular points, otherwise 
	$\xi$ is called a {\em critical value}. Please note that,
	according to this definition, any $\xi \in \M_2$ that is 
	not in the image of $F$ is regular. 
	The well-known ({\em Morse\/}-){\em Sard theorem\/} (see, e.g., 
	Hirsch \cite{Hi}, p.69) says that the set of regular values 
	of $F$ is residual (i.e., it contains the intersection of 
	countably many sets that are open and dense in $\M_2$) and thus 
	dense in $\M_2$ and the critical values of $F$ make up a set 
	of measure zero in $\M_2$.

	For the lens map $f_p : \D_p \longrightarrow \N$, we call 
	the set 
\begin{equation}\label{eq:caustic}
	{\mathrm{Caust}}(f_p) = \big\{ \, \xi \in \N \, \big| 
	\, \xi {\text{ is a critical value of }} f_p \, \big\}
\end{equation}
	the {\em caustic\/} of $f_p$. The Sard theorem then
	implies the following result.
\begin{proposition}\label{prop:sard}
	The caustic ${\mathrm{Caust}}(f_p)$ is a set 
	of measure zero in $\N$ and its 
	complement $\N \setminus {\mathrm{Caust}}(f_p)$ is residual and 
	thus dense in $\N$.
\end{proposition}
	Please note that ${\mathrm{Caust}}(f_p)$ need not be closed 
	in $\N$. Counter-examples can be constructed easily by starting 
	with situations where the caustic is closed and then excising 
	points from spacetime. For lens maps defined on the whole 
	celestial sphere, however, we have the following result.
\begin{proposition}\label{prop:ccompact}
	If $\D_p = \s_p$, the caustic ${\mathrm{Caust}}(f_p)$ is
	compact in $\N$.
\end{proposition}
	This is an obvious consequence of the fact that $\s_p$ is 
	compact and that $f_p$ and its first derivative are continuous.
	
	As the domain and the target space of $f_p$ have the
	same dimension, $Y_p \in \D_p$ is a regular point of $f_p$
	if and only if the differential $T_{Y_p}f_p : 
	T_{Y_p} \s_p\longrightarrow T_{f_p(Y_p)} \N$ is an 
	isomorphism. In this case $f_p$ maps a neighborhood of $Y_p$ 
	diffeomorphically onto a neighborhood of $f_p(Y_p)$. The 
	differential $T_{Y_p}f_p$ may be either orientation-preserving 
	or orientation-reversing. To make this notion precise we have
	to choose an orientation for $\s_p$ and an orientation for 
	$\N$. For the celestial sphere $\s_p$ it is natural to choose
	the orientation according to which the origin of the
	tangent space $T_p \M$ is to the inner side of $\s_p$. The
	target manifold $\N$ is orientable by assumption, but in general
	there is no natural choice for the orientation. Clearly, choosing
	an orientation for $\N$ fixes an orientation for $\T$, because
	the vector field $W$ gives us an orientation for the fibers.
	We shall say that the orientation of $\N$ is {\em adapted\/} 
	to some point $Y_p \in \D_p$ if the geodesic with initial vector 
	$Y_p$ meets $\T$ at the inner side. If $\D_p$ is connected, the 
	orientation of $\N$ that is adapted to some $Y_p \in \D_p$ 
	is automatically adapted to all other elements of
	$\D_p$. Using this terminology, we may now introduce the
	following definition.
\begin{definition}\label{def:parity}
	A regular point $Y_p \in \D_p$ of the lens map $f_p$ is 
	said to have {\em even parity\/} (or {\em odd parity}, 
	respectively) if $T_{Y_p}f_p$ is orientation-preserving 
	(or orientation-reversing, respectively) with respect to the 
	natural orientation on $\s_p$ and the orientation adapted to 
	$Y_p$ on $\N$. For a regular value $\xi \in \N$ of 
	the lens map, we denote by $n_+ (\xi )$ (or $n_- (\xi )$,
	respectively) the number of elements in $f_p^{-1}(\xi )$ with 
	even parity (or odd parity, respectively).  
\end{definition}
	Please note that $n_+(\xi)$ and $n_-(\xi)$ may be infinite,
	see the Schwarzschild example in Subsection~\ref{subsec:string}
	below. A criterion for $n_{\pm}(\xi)$ to be finite will be 
	given in Proposition~\ref{prop:finite} below.

	Definition~\ref{def:parity} is relevant for gravitational
	lensing in the following sense. The assumption that $Y_p$ 
	is a regular point of $f_p$ implies that an observer at $p$ 
	sees a neighborhood of $\xi = f_p(Y_p)$ in $\N$ as a 
	neighborhood of $Y_p$ at his or her celestial sphere.
	If we compare the case that $Y_p$ has odd parity with the case
	that $Y_p$ has even parity, then the appearance of the
	neighborhood in the first case is the mirror image of its
	appearance in the second case. This difference is observable 
	for a light source that is surrounded by some irregularly 
	shaped structure, e.g. a galaxy with curved jets or with
	lobes. 

	If $\xi$ is a regular value of $f_p$, it is obvious that 
	the points in $f_p^{-1} (\xi )$ are isolated, i.e., any 
	$Y_p$ in $f_p^{-1} (\xi )$ has a neighborhood in $\D_p$ that
	contains no other point in $f_p^{-1} (\xi )$. This follows 
	immediately from the fact that $f_p$ maps a neighborhood 
	of $Y_p$ diffeomorphically onto its image. In the next 
	section we shall formulate additional assumptions such 
	that the set $f_p^{-1} ( \xi )$ is finite, i.e., such that
	the numbers $n_{\pm} (\xi)$ introduced in 
	Definition~\ref{def:parity} are finite. It is the main purpose 
	of the next section to demonstrate that then the difference 
	$n_+ (\xi ) - n_- ( \xi )$ has some topological invariance 
	properties. As a preparation for that we notice the following 
	result which is an immediate consequence of the fact that 
	the lens map is a local diffeomorphism near each regular point.
\begin{proposition}\label{prop:n}
	$n_+$ and $n_-$ are constant on each connected 
	component of $f_p (\D_p) \setminus {\mathrm{Caust}}(f_p)$.
\end{proposition}
	Hence, along any continuous curve in $f_p(\D_p)$ that 
	does not meet the caustic of the lens map, the numbers $n_+$ 
	and $n_-$ remain constant, i.e., the observer at $p$ sees the same 
	number of images for all light sources on this curve. If a 
	curve intersects the caustic, the number of images will jump. 
	In the next section we shall prove that $n_+$ and $n_-$ always 
	jump by the same amount (under conditions making sure that 
	these numbers are finite), i.e., the total number of images 
	always jumps by an even number. This is well known in the 
	quasi-Newtonian approximation formalism, see, e.g., Schneider,
	Ehlers and Falco \cite{SEF}, Section~6. 

	If ${\mathrm{Caust}}(f_p)$ is empty, transversality guarantees
	that $f_p(\D_p)$ is open in $\N$ and, thus a manifold.
	Proposition~\ref{prop:n} implies that, in this case,
	$f_p$ gives a $C^{\infty}$ covering map from $\D_p$ onto
	$f_p(\D_p)$. As a $C^{\infty}$ covering map onto a simply
	connected manifold must be a global diffeomorphism, this
	implies the following result.
\begin{proposition}\label{prop:cover}
	Assume that ${\mathrm{Caust}}(f_p)$ is empty and that $f_p(\D_p)$
	is simply connected. Then $f_p$ gives a global diffeomorphism
	from $\D_p$ onto $f_p(\D_p)$.
\end{proposition}
	In other words, the formation of a caustic is necessary for 
	multiple imaging provided that $f_p(\D_p)$ is simply connected.
	In Subsection~\ref{subsec:string} below we shall consider the
	spacetime of a non-transparent string. This will demonstrate
	that the conclusion of Proposition~\ref{prop:cover} is not true
	without the assumption of $f_p(\D_p)$ being simply connected.
	
	In the rest of this subsection we want to relate the caustic
	of the lens map to the caustic of the past light cone of $p$.
	The past light cone of $p$ can be defined as the image set 
	in $\M$ of the map 
\begin{equation}\label{eq:Fp}
	F_p : (s, Y_p) \longmapsto {\mathrm{exp}}_p (sY_p)
\end{equation}
	considered on its maximal domain in 
	$]\, 0 \, , \, \infty \, [ \: \times \, \s_p \,$, 
	and its caustic can be defined as the set of critical values 
	of $F_p$. In other words, $q \in \M$ is in the caustic of 
	the past light cone of $p$ if and only if there is an 
	$s_0 \in \; ]\, 0 \, , \, \infty\, [ \,$ and a $Y_p \in \s_p$ 
	such that the differential $T_{(s_0,Y_p)} F_p$ has rank 
	$k < 3$. In that case one says that the point 
	$q = {\mathrm{exp}}_p(s_0 Y_p)$ is {\em conjugate\/} to 
	$p$ along the geodesic $s \longmapsto {\mathrm{exp}}_p(sY_p)$, 
	and one calls the number $m = 3-k$ the {\em multiplicity\/} 
	of this conjugate point. As $F_p ( \, \cdot\,  , Y_p)$ is 
	always an immersion, the multiplicity can take the values 
	1 and 2 only. (This formulation is equivalent to the 
	definition of conjugate points and their multiplicities in 
	terms of {\em Jacobi vector fields\/} which may be more 
	familiar to the reader.) It is well known, but far from
	trivial, that along every lightlike geodesic conjugate 
	points are isolated. Hence, in a compact parameter interval
	there are only finitely many points that are conjugate to 
	a fixed point $p$. A proof can be found, e.g., in Beem, 
	Ehrlich and Easley \cite{BEE}, Theorem~10.77. 

	After these preparations we are now ready to establish
	the following proposition. We use the notation introduced
	in Definition~\ref{def:lensmap}.
\begin{proposition}\label{prop:caustic}
	An element $Y_p \in \D_p$ is a regular point of the lens map 
	if and only if the point ${\mathrm{exp}}_p(w_p(Y_p)Y_p)$ is 
	not conjugate to $p$ along the geodesic $s \longmapsto 
	{\mathrm{exp}}_p (sY_p)$. A regular point $Y_p \in \D_p$ has 
	even parity $($or odd parity, respectively$\, )$ if and only 
	if the number of points conjugate to $p$ along the geodesic 
	$[\, 0 \, ,w_p(Y_p)] \; \longrightarrow \M \, , \: s 
	\longmapsto {\mathrm{exp}}_p (sY_p)$ is even $($or odd, 
	respectively$\, )$. Here each conjugate point is to be 
	counted with its multiplicity.
\end{proposition}
\begin{proof}
	In terms of the function (\ref{eq:Fp}), the lens map can be 
	written in the form 
\begin{equation}\label{eq:fF}
	f_p(Y_p) = \pi _W \big( F_p ( w_p (Y_p), Y_p ) \big) \; .
\end{equation}
	As $s \longmapsto F_p (s, Y_p)$ is an immersion transverse 
	to $\T$ at $s = w_p(Y_p)$ and $\pi _W$ is a submersion, the 
	differential of $f_p$ at $Y_p$ has rank 2 if and only if the 
	differential of $F_p$ at $(w_p(Y_p),Y_p)$ has rank 3. This 
	proves the first claim. For proving the second claim define, for 
	each $s \in [0,w_p(Y_p)]$, a map 
\begin{equation}\label{eq:Phi}
	\Phi_s : T_{Y_p} \s_p \longrightarrow T_{f_p(Y_p)}\N
\end{equation}
	by applying to each vector in $T_{Y_p} \s_p$ the differential
	$T_{(s,Y_p)} F_p$, parallel-transporting the result along the
	geodesic $F_p(\, \cdot \, , Y_p )$ to the point $q = F_p\big(
	w_p(Y_p), Y_p \big)$ and then projecting down to $T_{f_p(Y_p)} \N$. 
	In the last step one uses the fact that, by transversality, any
	vector in $T_q \M$ can be uniquely decomposed into a vector
	tangent to $\T$ and a vector tangent to the geodesic $F_p(\, 
	\cdot \, , Y_p)$. For  $s = 1$, this map $\Phi_s$ gives the 
	differential of the lens map. We now choose a basis in 
	$T_{Y_p} \s_p$ and a basis in $T_{f_p(Y_p)} \N$, thereby
	representing the map $\Phi_s$ as a $(2 \times 2)$-matrix. 
	We choose the first basis right-handed with respect to the 
	natural orientation on $\s_p$ and the second basis right-handed
	with respect to the orientation on $\N$ that is adapted to $Y_p$.
	Then ${\mathrm{det}}(\Phi_0)$ is positive as the parallel
	transport gives an orientation-preserving isomorphism. 
	The function $s \longmapsto {\mathrm{det}}(\Phi_s)$ has 
	a single zero whenever $F_p(s,Y_p)$ is a conjugate point of 
	multiplicity one and it has a double zero whenever $F_p(s,Y_p)$ 
	is a conjugate point of multiplicity two. Hence, the sign
	of ${\mathrm{det}}(\Phi_1)$ can be determined by counting
	the conjugate points.
\end{proof}

	This result implies that $\xi \in \N$ is a regular value
	of the lens map $f_p$ whenever the worldline $\xi$ does not
	pass through the caustic of the past light cone of $p$. The
	relation between parity and the number of conjugate
	points is geometrically rather evident because each 
	conjugate point is associated with a ``crossover'' of
	infinitesimally neighboring light rays.
	
%%%%%%%%%%%%%%%%%%%%%%%%%%%%%%%%%%%%%%%%%%%%%%%%%%%%%%%%%%%%%%%%%%%%%%%%%%%%%%

\section{The mapping degree of the lens map}\label{sec:degree}
	The mapping degree (also known as Brouwer degree) is one of the
	most powerful tools in differential topology. In this section
	we want to investigate what kind of information could be gained
	from the mapping degree of the lens map, providing it can be 
	defined. 

	For the reader's convenience we briefly summarize definition 
	and main properties of the mapping degree, following closely 
	Choquet-Bruhat, Dewitt-Morette, and Dillard-Bleick \cite{CDD}, 
	pp.477. For a more abstract approach, using homology theory, 
	the reader may consult Dold \cite{Do}, Spanier \cite{Sp} or
	Bredon \cite{Br}. In this article we shall not use homology
	theory with the exception of the proof of 
	Proposition~\ref{prop:sln}.

	The definition of the mapping degree is based on the following
	observation.
\begin{proposition}\label{prop:F}
	Let $F : {\overline{\D}} \subseteq \M_1 \longrightarrow \M_2$ 
	be a continuous map, where $\M_1$ and $\M_2$ are oriented 
	connected manifolds of the same dimension, $\D$ is an open subset
	of $\M_1$ with compact closure ${\overline{\D}}$ and $F|_{\D}$
	is a $C^{\infty}$ map. $($Actually, $C^1$ would do.$)$ Then for 
	every $\xi \in \M_2 \setminus F(\partial \D)$ which is a 
	regular value of $F|_{\D}$, the set $F^{-1}(\xi)$ is finite.
\end{proposition}
\begin{proof} 
	By contradiction, let us assume that there is a 
	sequence $(y_i)_{i \in \n }$ with pairwise different elements in 
	$F^{-1} ( \xi )$. By compactness of ${\overline{\D}}$,
	we can choose an infinite subsequence that converges
	towards some point $y_{\infty} \in {\overline{\D}}$.
	By continuity of $F$, $F (y_{\infty}) = \xi$, so the hypotheses
	of the proposition imply that $y_{\infty} \notin \partial \D$.
	As a consequence, $y_{\infty}$ is a regular point of $F|_{\D}$, 
	so it must have an open neighborhood in $\D$ that does not 
	contain any other element of $F^{-1} ( \xi )$. This contradicts 
	the fact that a subsequence of $(y_i)_{i \in {\mathbb{N}}}$ 
	converges towards $y_{\infty}$.
\end{proof}
	If we have a map $F$ that satisfies the hypotheses of 
	Proposition~\ref{prop:F}, we can thus define, for every $\xi \in \M_2 
	\setminus F (\partial \D)$ which is a regular value of 
	$F|_{\D}$, 
\begin{equation}\label{eq:deg}
	\g (F, \xi) = \sum_{y \, \in \, F^{-1} (\xi )} {\mathrm{sgn}} (y) \: ,
\end{equation}
	where sgn($y$) is defined to be $+1$ if the differential
	$T_yF$ preserves orientation and $-1$ if $T_yF$ reverses 
	orientation. If $F^{-1} (\xi )$ is the empty set, the
	right-hand side of (\ref{eq:deg}) is set equal to zero.
	The number $\g (F, \xi )$ is called the {\em mapping degree\/}
	of $F$ at $\xi$. Roughly speaking, $\g (F, \xi )$ tells how often
	the image of $F$ covers the point $\xi$, counting each ``layer''
	positive or negative depending on orientation. 

	The mapping degree has the following properties (for proofs
	see Choquet-Bruhat, Dewitt-Morette, and Dillard-Bleick 
	\cite{CDD}, pp.477).
\\[0.1cm]
	{\bf Property~A}: $\g (F, \xi ) = \g (F, \xi ' )$ whenever 
	$\xi $ and $\xi '$ are in the same connected component
	of $\M_2 \setminus F(\partial \D)$.
\\[0.1cm]
	{\bf Property~B}: $\g (F, \xi ) = \g (F', \xi )$ whenever
	$F$ and $F'$ are homotopic, i.e., whenever there is a
	continuous map $\Phi : [0,1] \times {\overline{\D}} \longrightarrow
	\M_2\, , \: (s,y) \longmapsto \Phi_s(y)$ with $\Phi_0 = F$ 
	and $\Phi_1 = F'$ such that $\g (\Phi _s , \xi )$ is defined
	for all $s \in [0,1]$.

\vspace{0.1cm}
 	Property~A can be used to extend the definition of 
	$\g (F, \xi )$ to the non-regular values $\xi \in \M_2 
	\setminus F(\partial \D)$. Given the fact that, by the
	Sard theorem, the regular values are dense in 
	$\M_2$, this can be done just by continuous extension.	
	
	Property~B can be used to extend the definition of 
	$\g (F, \xi )$ to continuous maps $F : {\overline{D}} 
	\longrightarrow \M_2$ which are not necessarily 
	differentiable on $\D$. Given the fact that the 
	$C^{\infty}$ maps are dense in the continuous maps
	with respect to the $C^0$-topology, this can be done
	again just by continuous extension.

	We now apply these general results to the lens map
	$f_p : \D_p \longrightarrow \N$. In the case $\D_p \neq
	\s_p$ it is necessary to extend the domain of the lens
	map onto a compact set to define the degree of the lens
	map. We introduce the following definition.
\begin{definition}\label{def:extend}
	A map ${\overline{f_p}} : {\overline{\D_p}} \subseteq \M_1 
	\longrightarrow  \M_2$ is called an {\em extension\/} of 
	the lens map $f_p : \D_p \longrightarrow \N$ if 
\\
	(a) $\M_1$ is an orientable manifold that contains $\D_p$ 
	as an open submanifold;
\\
	(b) $\M_2$ is an orientable manifold that contains $\N$
	as an open submanifold;
\\
	(c) the closure ${\overline{\D_p}}$ of $\D_p$ in $\M_1$
	is compact;
\\
	(d) ${\overline{f_p}}$ is continuous and the restriction
	of ${\overline{f_p}}$ to $\D_p$ is equal to $f_p$.
\end{definition}
	If the lens map is defined on the whole celestial sphere,
	$\D_p = \s_p$, then the lens map is an extension of 
	itself, ${\overline{f_p}} = f_p$, with $\M_1 = \s_p$ 
	and $\M_2 = \N$. If $\D_p \neq \s_p$, one
	may try to continuously extend $f_p$ onto the closure 
	of $\D_p$ in $\s_p$, thereby getting an extension
	with $\M_1 = \s_p$ and $\M_2 = \N$. If this does not work, 
	one may try to find some other extension. The string spacetime in
	Subsection~\ref{subsec:string} below will provide us with
	an example where an extension exists although
	$f_p$ cannot be continuously extended from $\D_p$ onto 
	its closure in $\s_p$. The spacetime around a spherically
	symmetric body with $R_o<3m$ will provide us with an example
	where the lens map admits no extension at all, see
	Subsection~\ref{subsec:string} below.

	Applying Proposition~\ref{prop:F} to the case $F = 
	{\overline{f_p}}$ immediately gives the following result.
\begin{proposition}\label{prop:finite}
	If the lens map $f_p : \D_p \longrightarrow \N$ admits an
	extension ${\overline{f_p}} : {\overline{\D_p}} \subseteq \M_1
	\longrightarrow \M_2$, then for all regular values $\xi \in
	\N \setminus {\overline{f_p}} (\partial \D_p)$ the set 
	$f_p^{-1} (\xi)$ is finite, so the numbers $n_+(\xi)$ and
	$n_-(\xi)$ introduced in Definition~$\ref{def:parity}$ are
	finite.
\end{proposition}
	If ${\overline{f_p}}$ is an extension of the lens map $f_p$,
	the number $\g ({\overline{f_p}}, \xi )$ is a well defined
	integer for all $\xi \in \N \setminus {\overline{f_p}}(
	\partial \D_p )$, provided that we have chosen an orientation
	on $\M_1$ and on $\M_2$. The number $\g({\overline{f_p}},
	\xi)$ changes sign if we change the orientation on $\M_1$ 
	or on $\M_2$. This sign ambiguity can be removed if $\D_p$ 
	is connected. Then we know from the preceding section that 
	$\N$ admits an orientation that is adapted to all $Y_p \in 
	\D_p$. As $\N$ is connected, this determines an orientation 
	for $\M _2$. Moreover, the natural orientation on $\s_p$ 
	induces an orientation on $\D_p$ which, for $\D_p$ connected, 
	gives an orientation for $\M _1$. 

	In the rest of this paper we shall only be concerned with the 
	situation that $\D _p$ is connected, and we shall always tacitly 
	assume that the orientations have been chosen as indicated above, 			thereby fixing the sign of $\g (f_p)$. Now comparison of 
	(\ref{eq:deg}) with Definition~\ref{def:parity} shows that
\begin{equation}\label{eq:degn}
	\g ({\overline{f_p}} , \xi ) = n_+ (\xi ) - n_- (\xi ) 
\end{equation}
	for all regular values in $\N \setminus {\overline{f_p}}
	(\partial \D_p)$. Owing to Property~A, this has the 
	following consequence.
\begin{proposition}\label{prop:nn}
	Assume that $\D_p$ is connected and that the lens map
	admits an extension ${\overline{f_p}} : 
	{\overline{\D_p}} \subseteq \M_1 \longrightarrow \M_2$. 
	Then $n_+ (\xi ) - n_- (\xi ) = n_+ (\xi' ) - n_- (\xi' )$
	for any two regular values $\xi$ and $\xi '$ which are
	in the same connected component of $\N \setminus 
	{\overline{f_p}} (\partial \D_p)$. In particular, 
	$n_+ (\xi ) + n_- (\xi )$ is odd if and only if 
	$n_+ (\xi' ) + n_- (\xi' )$ is odd. 
\end{proposition}
	We know already from Proposition~\ref{prop:n} that the 
	numbers $n_+$ and $n_-$ remain constant along each 
	continuous curve in $f_p(\D_p)$ that does not meet
	the caustic of $f_p$. Now let us consider a continuous curve 
	$\alpha \, : \; ]- \varepsilon _0 \, , \, \varepsilon _0 \, [
	\; \longrightarrow f_p (\D_p)$ that meets the caustic at 
	$\alpha (0)$ whereas $\alpha (\varepsilon )$ is a 
	regular value of $f_p$ for all $\varepsilon \neq 0$. 
	Under the additional assumptions that $\D_p$ is 
	connected, that $f_p$ admits an extension, and 
	that $\alpha (0) \notin {\overline{f_p}} (\partial \D_p )$, 
	Proposition~\ref{prop:nn} tells us that $n_+ \big( \alpha 
	(\varepsilon) \big) - n_- \big( \alpha ( \varepsilon ) \big)$ 
	remains constant when $\varepsilon$ passes through zero. In 
	other words, $n_+$ and $n_-$ are allowed to jump only by 
	the same amount. As a consequence, the total number of 
	images $n_+ + n_-$ is allowed to jump only by an even 
	number.
	
	We now specialize to the case that the lens
	map is defined on the whole celestial sphere, $\D_p = \s_p$.
	Then the assumption of $f_p$ admitting an extension is 
	trivially satisfied, with ${\overline{f_p}} = f_p$, and the 
	degree $\g (f_p , \xi )$ is a well-defined integer 
	for all $\xi \in \N$. Moreover, $\g (f_p , \xi )$ is a 
	constant with respect to $\xi$, owing to Property~A. It 
	is then usual to write simply $\g (f_p)$ instead of $\g (f_p ,
	\xi )$. Using this notation, (\ref{eq:degn}) simplifies to
\begin{equation}\label{eq:degng}
	\g (f_p ) = n_+ (\xi ) - n_- (\xi ) 
\end{equation}
	for all regular values $\xi$ of $f_p$.
	Thus, the total number of images 
\begin{equation}\label{eq:gnum}
	n_+ (\xi ) + n_- (\xi ) = \g (f_p ) + 2  n_- (\xi )
\end{equation}
	is either even for all regular values $\xi$ or odd for all 
	regular values $\xi$, depending on whether $\g (f_p)$ is 
	even or odd. 

	In some gravitational lensing situations it 
	might be possible to show that there is one light source 
	$\xi \in \N$ for which $f_p^{-1}(\xi)$ consists of exactly 
	one point, i.e., $\xi$ is not multiply imaged. This situation
	is characterized by the following proposition.
\begin{proposition}\label{prop:unique}
	Assume that $\D_p = \s_p$ and that there is a regular value
	$\xi$ of $f_p$ such that $f_p^{-1}( \xi )$ is a single point.
	Then $|\g (f_p)| = 1$. In particular, $f_p$ must be 
	surjective and $\N$ must be diffeomorphic to the sphere 
	$S^2$.
\end{proposition}
\begin{proof}
	The result $|\g (f_p)| = 1$ can be read directly from
	(\ref{eq:degng}), choosing the regular value $\xi$ which
	has exactly one pre-image point under $f_p$. This implies that
	$f_p$ must be surjective since a non-surjective map has degree
	zero. So $\N$ being the continuous image of the compact set
	$\s_p$ under the continuous map $f_p$ must be compact. It is
	well known (see, e.g., Hirsch \cite{Hi}, p.130, Exercise 5)
	that for $n \ge 2$ the existence of a continuous map $F : S^n
	\longrightarrow \M_2$ with $\g(F) = 1$ onto a compact oriented
	$n$-manifold $\M_2$ implies that $\M_2$ must be simply
	connected. As the lens map gives us such a map onto $\N$ 
	(after changing the orientation of $\N$, if necessary), we 
	have thus found that $\N$ must be simply connected. Owing to 
	the well-known classification theorem of compact orientable 				two-dimensional manifolds (see, e.g., Hirsch \cite{Hi}, 
	Chapter 9), this implies that $\N$ must be diffeomorphic to 
	the sphere $S^2$.
\end{proof}

	In the situation of Proposition~\ref{prop:unique} we have
	$n_+(\xi ) + n_- (\xi) = 2 n_- (\xi) \pm 1$, for all $\xi \in
	\N \setminus {\mathrm{Caust}} (f_p)$, i.e., the total number
	of images is odd for all light sources $\xi \in \N \simeq S^2$
	that lie not on the caustic of $f_p$. The idea to use the 
	mapping degree for proving an odd number theorem in this way
	was published apparently for the first time in the introduction
	of McKenzie \cite{McK}. In Proposition~\ref{prop:unique} one 
	would, of course, like to drop the rather restrictive assumption
	that $f_p^{-1} (\xi)$ is a single point for some $\xi$. In the
	next section we consider a special situation where the result
	$| \g (f_p) | = 1$ can be derived without this assumption.

%%%%%%%%%%%%%%%%%%%%%%%%%%%%%%%%%%%%%%%%%%%%%%%%%%%%%%%%%%%%%%%%%%%%%%%%%%%%

\section{Simple lensing neighborhoods}\label{sec:sln}

	In this section we investigate a special class of spacetime
	regions that will be called ``simple lensing neighborhoods''. 
	Although the assumption of having a simple lensing neighborhood 
	is certainly rather special, we shall demonstrate in 
	Section~\ref{sec:examples} below that sufficiently many 
	examples of physical interest exist. We define simple lensing
	neighborhoods in the following way.
\begin{definition}\label{def:sln}
	$(\U ,\T , W)$ is called a {\em simple lensing neighborhood\/} in a 
	spacetime $(\M,g)$ if 
\\
	(a) $\U$ is an open connected subset of $\M$ and $\T$ is the boundary
	of $\U$ in $\M$;
\\
	(b) $(\, \T = \partial U , \, W \, )$ is a source surface in the 
	sense of Definition~\ref{def:TW};
\\
	(c) for all $p \in \U$, the lens map $f_p : \D_p \longrightarrow
	\N = \partial \U /W$ is defined on the whole celestial sphere,
	$\D_p = \s_p$;
\\
	(d) $\U$ does not contain an almost periodic lightlike geodesic.
\end{definition}
	Here the notion of being ``almost periodic'' is defined in the
	following way. Any immersed curve $\lambda : I \longrightarrow
	\U$, defined on a real interval $I$, induces a curve 
	${\hat{\lambda}} : I \longrightarrow P\U$ in
	the projective tangent bundle $P \U$ over $\U$ which is defined
	by ${\hat{\lambda}} (s) = \{ \, c {\dot{\lambda}} (s) \: | 
	\: c \in \R \: \}$. The curve $\lambda$ is called {\em almost 
	periodic\/} if there is a strictly monotonous sequence of 
	parameter values $(s_i)_{i \in \n}$ such that the sequence 
	$\big( {\hat{\lambda}} (s_i) \big) {}_{i \in \n}$ has an 
	accumulation point in $P \U$. Please note that Condition (d) 
	of Definition~\ref{def:sln} is certainly true if the strong 
	causality condition holds everywhere on $\U$, i.e., if there are
	no closed or almost closed causal curves in $\U$. Also, Condition 
	(d) is certainly true if every future-inextendible lightlike 
	geodesic in $\U$ has a future end-point in $\M$.

	Condition (d) should be viewed as adding a fairly mild
	assumption on the future-behavior of lightlike geodesics
	to the fairly strong assumptions on their past-behavior that 
	are contained in Condition (c). In particular,  
	Condition (c) excludes the possibility that 
	past-oriented lightlike geodesics are blocked or 
	trapped inside $\U$, i.e., it excludes 
	the case that $\U$ contains non-transparent deflectors. 
	Condition (c) requires, in addition, that the past-pointing
	lightlike geodesics are transverse to $\partial \U$ when
	leaving $\U$.  

	In the situation of a simple lensing neighborhood, we have
	for each $p \in \U$ a lens map that is defined on the whole
	celestial sphere, $f_p: \s_p \longrightarrow \N = \partial \U /W$.
	We have, thus, equation (\ref{eq:degng}) at our disposal which 
	relates the numbers $n_+(\xi)$ and $n_-(\xi)$, for any regular 
	value $\xi \in \N$, to the mapping degree of $f_p$. (Please recall 
	that, by Proposition~\ref{prop:finite}, $n_+(\xi)$ and $n_-(\xi)$ 
	are finite.) It is our main goal to prove that, in a simple lensing
	neighborhood, the mapping degree of the lens map equals $\pm 1$, so
	$n( \xi ) = n_+ ( \xi ) + n_- ( \xi )$ is odd for all regular
	values $\xi$. Also, we shall prove that a simple lensing neighborhood
	must be contractible and that its boundary must be diffeomorphic
	to $S^2 \times \R$. The latter result reflects the fact that the
	notion of simple lensing neighborhoods generalizes the notion of 
	asymptotically simple and empty spacetimes, with $\partial \U$
	corresponding to past lightlike infinity $\J ^-$, as will be detailed 
	in Subsection~\ref{subsec:asy} below. When proving the desired 
	properties of simple lensing neighborhoods we may therefore use
	several techniques that have been successfully applied to 
	asymptotically simple and empty spacetimes before.

	As a preparation we need the following lemma.
\begin{lemma}\label{lem:sln}
	Let $(\U ,\T ,W)$ be a simple lensing neighborhood in a spacetime 
	$(\M,g)$. Then there is a diffeomorphism $\Psi$ from the sphere bundle
	$\s = \big\{ Y_p \in \s_p \, \big| \, p \in \U \, \big\}$ of 
	lightlike directions over $\U$ onto the space $T \N \times \R ^2$
	such that the following diagramm commutes.
\begin{equation}\label{eq:cd}
\begin{matrix}
	&\; \s  & {\overset{\Psi}{\longrightarrow}} & T\N \times \R ^2 
\\[0.2cm]
	& i_p \uparrow \; \; & \, &  \; \; \downarrow {\mathrm{pr}} 
\\[0.2cm]
	& \; \s _p & {\overset{f_p}{\longrightarrow}}  & \N \;
\end{matrix}
\end{equation}
	Here $i_p$ denotes the inclusion map and ${\mathrm{pr}}$ is defined 			by dropping the second factor and projecting to the foot-point. 
\end{lemma}
\begin{proof} 
	We fix a trivialization for the bundle $\pi _W : \T \longrightarrow
	\N$ and identify $\T$ with $\N \times \R$. Then we consider the 
	bundle $ \B = \big\{ X_q \in \B _q \, \big| \, q \in 
	\T \, \big\}$ over $\T$, where $\B_q \subset \s_q$ is, by 
	definition, the subspace of all lightlike directions that are 
	tangent to past-oriented lightlike geodesics that leave $\U$
	transversely at $q$. Now we choose for each $q \in \T$ a vector
	$Q_q \in T_q \M$, smoothly depending on $q$, which is non-tangent
	to $\T$ and outward pointing. With the help of this vector field $Q$
	we may identify $\B$ and $ T \N \times \R $ as bundles over 
	$\T \simeq \N \times \R$ in the following way. Fix $\xi \in \N$, 
	$X_{\xi} \in T _{\xi} \N$ and $s \in \R$ and view the tangent space 
	$T_{\xi} \N$ as a natural sub\-space of $T_q (\N \times \R)$, where 
	$q = (\xi , s)$. Then the desired identification is given by 
	associating the pair $(X_{\xi},s)$ with the direction spanned by 
	$Z_q = X_{\xi} + Q_q - \alpha \, W(q)$, where the number $\alpha$ is 
	uniquely determined by the requirement that $Z_q$ should be lightlike 
	and past-pointing. -- Now we consider the map 
\begin{equation}\label{eq:pi}
	\pi : \s \longrightarrow \B \simeq T \N \times \R
\end{equation}
	given by following each lightlike geodesic from a point $p \in \U$
	into the past until it reaches $\T$, and assigning the tangent 
	direction at the end-point to the tangent direction at the 
	initial point. As a matter of fact, (\ref{eq:pi}) gives a 
	principal fiber bundle with structure group $\R$. To prove 
	this, we first observe that the geodesic spray induces a vector 
	field without zeros on $\s$. By multiplying this vector field 
	with an appropriate function we get a vector field whose flow 
	is defined on all of $\R \times \s$ (see the second paragraph after 
	Definition \ref{def:TW} for how to find such a function). 
	The flow of this rescaled vector field defines an $\R$-action 
	on $\s$ such that (\ref{eq:pi}) can be identified with the 
	projection onto the space of orbits. Conditions (c) and (d) 
	of Definition~\ref{def:sln} guarantee that no orbit is closed 
	or almost closed. Owing to a general result of Palais
	\cite{Pa}, this is sufficient to prove that this action  
	makes (\ref{eq:pi}) into a principal fiber bundle with structure 
	group $\R$. However, any such bundle is trivializable, see, e.g.,
	Kobayashi and Nomizu \cite{KN}, p.57/58.
	Choosing a trivialization for (\ref{eq:pi}) gives us the
	desired diffeomorphism $\Psi$ from $\s$ to $\B \times \R \simeq
	T \N \times \R ^2$. The commutativity of the diagram (\ref{eq:cd})
	follows directly from the definition of the lens map $f_p$.
\end{proof}
	With the help of this lemma we will now prove the following 
	proposition which is at the center of this section.
\begin{proposition}\label{prop:sln}
	Let $(\U, \T , W)$ be a simple lensing neighborhood in a spacetime 
	$(\M,g)$. Then 
\\
	{\em (a)} $\N = \T /W$ is diffeomorphic to the standard 
	$2$-sphere $S^2$;
\\
	{\em (b)} $\U$ is contractible;
\\
	{\em (c)} for all $p \in \U$, the lens map $f_p : \s_p \simeq S^2
	\longrightarrow \N \simeq S^2$ has $ | \g (f_p) | = 1$; 
	in particular, $f_p$ is surjective.
\end{proposition}
\begin{proof}
	In the proof of part (a) and (b) we shall adapt techniques
	used by Newman and Clarke \cite{NC,Ne} in their study of
	asymptotically simple and empty spacetimes. To that end
	it will be necessary to assume that the reader is familiar 
	with homology theory. With the sphere bundle $\s$, introduced in
	Lemma~\ref{lem:sln}, we may associate the {\em Gysin homology 
	sequence\/} 
\begin{equation}\label{eq:Gysin}
	{} \dots {} \longrightarrow H_m (\s ) \longrightarrow H_m (\U ) \longrightarrow
	H_{m-3} (\U ) \longrightarrow H_{m-1} ( \s ) \longrightarrow \dots
\end{equation}
	where $H_m( {\mathcal{X}} )$ denotes the $m^{\mathrm{th}}$ 
	homology group of the space ${\mathcal{X}}$ with coefficients 
	in a field $\F$. For any choice of $\F$, the Gysin sequence is 
	an exact sequence of abelian groups, see, e.g., Spanier 
	\cite{Sp}, p.260 or, for the analogous sequence of 
	cohomology groups, Bredon \cite{Br}, p.390. By 
	Lemma~\ref{lem:sln}, $\s$ and $\N$ have the same homotopy type, 
	so $H_m ( \s )$ and $H_m (\N )$ are isomorphic. 
	Upon inserting this into (\ref{eq:Gysin}), we use the fact 
	that $H_m ( \U ) = {\boldsymbol{1}} \, ( \, = \,$trivial 
	group consisting of the unit element only) for $m > 4$ and 
	$H_m (\N ) = {\boldsymbol{1}}$ for $m > 2$ because 
	${\mathrm{dim}} (\U ) = 4$ and ${\mathrm{dim}} (\N ) = 2$. 
	Also, we know that $H_0 (\U ) = \F$ and $H_0 (\N ) = \F$ 
	since $\U$ and $\N$ are connected. Then the exactness of 
	the Gysin sequence implies that 
\begin{equation}\label{eq:Hms}
	H_m (\U ) = {\boldsymbol{1}} \quad {\mathrm{for}} \; \; m > 0
\end{equation}
	and
\begin{equation}\label{eq:HmN}
	H_1 (\N ) = {\boldsymbol{1}} \; , \quad H_2 (\N ) = \F \, .
\end{equation}
	From (\ref{eq:HmN}) we read that $\N$ is compact since otherwise 
	$H_2( \N ) = {\boldsymbol{1}}$. Moreover, we observe that $\N$ has 
	the same homology groups and thus, in particular, the same Euler 
	characteristic as the 2-sphere. It is well known that any two compact 
	and orientable 2-manifolds are diffeomorphic if and only if they 
	have the same Euler characteristic (or, equivalently, the same genus), 
	see, e.g., Hirsch \cite{Hi}, Chapter 9. We have thus proven part 
	(a) of the proposition. -- To prove part (b) we consider the end of 
	the exact {\em homotopy sequence\/} of the fiber bundle $\s$ over 
	$\U$, see, e.g., Frankel \cite{Fr}, p.600,
\begin{equation}\label{eq:homseq}
	{} \dots {} \longrightarrow \pi _1 (\s )
	\longrightarrow \pi_1 (\U ) \longrightarrow {\boldsymbol{1}} \, .
\end{equation} 
	As $\s$ has the same homotopy type as $\N \simeq S^2$, we may
	replace $\pi_1 (\s )$ with $\pi_1 (S^2 ) = {\boldsymbol{1}}$, 
	so the exactness of (\ref{eq:homseq}) implies that $\pi _1 (\U ) =
	{\boldsymbol{1}}$, i.e., that $\U$ is simply connected. If, for some
	$m > 1$, the homotopy group $\pi _m (\U )$ would be different from 
	${\boldsymbol{1}}$, the Hurewicz isomorphism theorem (see, e.g., 
	Spanier \cite{Sp}, p.394 or Bredon \cite{Br}, p.479, Corollary 10.10.) 
	would give a contradiction to (\ref{eq:Hms}). Thus,
	$\pi _m (\U ) = {\boldsymbol{1}}$ for all $m \in \n$, i.e., 
	$\U$ is contractible. -- We now prove part (c). Since $\U$ is 
	contractible, the tangent bundle $T \U$ and thus the sphere bundle 
	$\s$ over $\U$ admits a global trivialization, 
	$ \s \simeq \U \times S^2$. Fixing such a 
	trivialization and choosing a contraction that collapses $\U$ 
	onto some point $p \in \U$ gives a contraction ${\tilde{i_p}} : 
	\s \longrightarrow \s_p \,$. Together with the inclusion map
	$i_p : \s_p \longrightarrow \s$	this gives us a homotopy 
	equivalence between $\s_p$ and $\s$. $($Please recall that a 
	{\em homotopy equivalence\/} between two topological spaces 
	${\mathcal{X}}$ and ${\mathcal{Y}}$ is a pair of continuous maps 
	$\varphi: {\mathcal{X}} \longrightarrow {\mathcal{Y}}$ and 
	${\tilde{\varphi}}: {\mathcal{Y}} \longrightarrow {\mathcal{X}}$ 
	such that $\varphi \circ {\tilde{\varphi}}$ can be continuously 
	deformed into the identity on ${\mathcal{Y}}$ and ${\tilde{\varphi}} 
	\circ \varphi$ can be continuously deformed into the identity on 
	${\mathcal{X}}.)$ On the other hand, the projection ${\mathrm{pr}}$ 
	from (\ref{eq:cd}), together with the zero section $\tilde{\mathrm{pr}} 
	: \N \longrightarrow T \N \times \R ^2$ gives a homotopy equivalence
	between $T \N \times \R ^2$ and $\N$. As a consequence, the
	diagram (\ref{eq:cd}) tells us that the lens map $f_p = {\mathrm{pr}}
	\circ \Psi \circ i_p$ together with the map ${\tilde{f}}{}_p
	= {\tilde{i}}{}_p \circ \Psi ^{-1} \circ {\tilde{\mathrm{pr}}}$
	gives a homotopy equivalence between $\s_p \simeq S^2$ and 
	$\N \simeq S^2$, so $f_p \circ {\tilde{f}}{}_p$ is homotopic to 
	the identity. Since the mapping degree is a homotopic invariant 
	(please recall Property~B of the mapping degree from 
	Section~\ref{sec:degree}), this implies that 
	$\g ( f_p \circ {\tilde{f}}{}_p ) = 1$. Now the product 
	theorem for the mapping degree (see, e.g., Choquet-Bruhat, 						Dewitt-Morette, and Dillard-Bleick \cite{CDD}, p.483)
	yields $\g (f_p) \, \g ({\tilde{f}}{}_p ) = 1$. As the 
	mapping degree is an integer, this can be true only if 
	$ \g (f_p )  = \g ( {\tilde{f}}{}_p ) = \pm 1$. In particular, 
	$f_p$ must be surjective since otherwise $\g (f_p ) = 0$.
\end{proof}

	In all simple examples to 
	which this proposition applies the degree of $f_p$ is, actually, 
	equal to $+1$, and it is hard to see whether examples with 
	$\g (f_p) = -1 $ do exist. The following consideration is 
	quite instructive. If we start with a simple lensing neighborhood
	in a flat spacetime (or, more generally, in a conformally flat 
	spacetime), then conjugate points cannot occur, so it is clear
	that the case $\g (f_p) = -1$ is impossible. If we now perturb
	the metric in such a way that the simple-lensing-neighborhood
	property is maintained during the perturbation, then, by Property~B 
	of the degree, the equation $\g (f_p ) = + 1$ is preserved.
	This demonstrates that the case $\g (f_p) = -1$ cannot occur
	for weak gravitational fields (or for small perturbations of
	conformally flat spacetimes such as Robertson-Walker spacetimes). 

	Among other things, Proposition~\ref{prop:sln} gives a good 
	physical motivation for studying degree-one maps from $S^2$ to 
	$S^2$. In particular, it is an interesting problem to characterize 
	the caustics of such maps. Please note that, by parts (a) and (c) 
	of Proposition~\ref{prop:sln}, $f_p(\D_p)$ is simply connected for 
	all $p \in \U$. Hence, Proposition~\ref{prop:cover} applies which 
	says that the formation of a caustic is necessary for multiple 
	imaging.  

	Owing to (\ref{eq:gnum}), part (c) of Proposition~\ref{prop:sln} 
	implies in particular that $n (\xi ) = n_+(\xi ) + n_-(\xi )$ is 
	odd for all worldlines of light sources $\xi \in \N$ that do not 
	pass through the caustic of the past light cone of $p$, i.e., if 
	only light rays within $\U$ are taken into account the observer
	at $p$ sees an odd number of images of such a worldline. It is 
	now our goal to prove a similar 'odd number theorem' 
	for a light source with worldline inside $\U$. As a preparation
	we establish the following lemma.
\begin{lemma}\label{lem:boundary}
	Let $(\U ,\T ,W)$ be a simple lensing neighborhood in a spacetime 
	$(\M,g)$ and $p \in \U$. Let $J^{-}(p, \U)$ denote, 
	as usual, the {\em causal past\/} of $p$ in $\U$, 
	i.e., the set of all points in $\M$ that can be reached from $p$ 
	along a past-pointing causal curve in $\U$. Let $\partial _{\U}
	J^-(p, \U )$ denote the boundary of $J^-(p, \U )$ in $\U$. Then
\\
	{\em (a)} every point $q \in \partial _{\U} J^-(p, \U )$ can be reached
	from $p$ along a past-pointing lightlike geodesic in $\U$;
\\
	{\em (b)} $\partial _{\U} J^-(p, \U )$ is relatively compact in $\M$.
\end{lemma}
\begin{proof}
	As usual, let $I^-(p, \U )$ denote the chronological past of
	$p$ in $\U$, i.e., the set of all points that can be 
	reached from $p$ along a past-pointing timelike curve in $\U$.
	To prove part (a), fix a point $q \in \partial _{\U} J^-(p, \U )$.
	Choose a sequence $(p_i)_{i \in \n}$ of points in $\U$ that converge
	towards $p$ in such a way that $p \in I^- (p_i , \U )$
	for all $i \in \n$. This implies that we can find for each 
	$i \in \n$ a past-pointing timelike curve $\lambda _i$
	from $p_i$ to $q$. Then the $\lambda _i$ are past-inextendible
	in $\U \setminus \{ q \}$. Owing to a standard lemma (see, e.g., 
	Wald \cite{Wa}, Lemma~8.1.5) this implies that the $\lambda _i$ 
	have a causal limit curve $\lambda$ through $p$ that is 						past-inextendible in $\U \setminus \{ q \}$. We want to show 
	that $\lambda$ is the desired lightlike geodesic. Assume that
	$\lambda$ is not a lightlike geodesic. Then $\lambda$ enters 
	into the open set $I^- (p, \U )$ (see Hawking and Ellis 
	\cite{HE}, Proposition~4.5.10), so $\lambda _i$ enters 
	into $I^- (p , \U )$ for $i$ sufficiently large. 
	This, however, is impossible since all $\lambda _i$ have 
	past end-point on $\partial _{\U} J^-(p, \U )$,
	so $\lambda$ must be a lightlike geodesic. It remains to show that
	$\lambda$ has past end-point at $q$. Assume that this is not true.
	Since $\lambda$ is past-inextendible in $\U \setminus \{ q \}$
	this assumption implies that $\lambda$ is past-inextendible 
	in $\U$, so by
	condition (c) of Definition~\ref{def:sln} $\lambda$ has past end-point
	on $\partial \U$ and meets $\partial \U$ transversely. As a 
	consequence, for $i$ sufficiently large $\lambda _i$ has to meet 					$\partial \U$ which gives a contradiction to the fact that all
	$\lambda _i$ are within $\U$. -- To prove part (b), we have to
	show that any sequence $(q_i)_{i \in \n}$ in 
	$\partial _{\U} J^-(p, \U )$ has an accumulation point in $\M$. 
	So let us choose such a sequence. From part (a) we know that there
	is a past-pointing lightlike geodesic $\mu _i$ from $p$
	to $q_i$ in $\U$ for all $i \in \n$. By compactness of $\s _p
	\simeq S^2$, the tangent directions to these geodesics at $p$
	have an accumulation point in $\s_p$. Let $\mu$ be the past-pointing
	lightlike geodesic from $p$ which is determined by this direction.
	By condition (c) of Definition~\ref{def:sln}, this geodesic 
	$\mu$ and each of the geodesics $\mu _i$ must have a past 
	end-point on $\partial \U$ if maximally extended inside $\U$. 
	We may choose an affine parametrization
	for each of those geodesics with the parameter ranging from the
	value 0 at $p$ to the value 1 at $\partial \U$. Then our sequence
	$(q_i)_{i \in \n}$ in $\U$ determines a sequence $(s_i)_{i \in \n}$
	in the interval $[0,1]$ by setting $q_i = \mu _i (s_i)$. By
	compactness of $[0,1]$, this sequence must have an accumulation
	point $s \in [0.1]$. This demonstrates that the $q_i$ must have
	an accumulation point in $\M$, namely the point $\mu (s)$.
\end{proof}
	We are now ready to prove the desired odd-number theorem for
	light sources with worldline in $\U$.
\begin{proposition}\label{prop:gamma}
	Let $(\U, \T ,W)$ be a simple lensing neighborhood in a spacetime 
	$(\M,g)$ and assume that $\U$ does not contain
	a closed timelike curve. Fix a point $p \in \U$ and a timelike 
	embedded $C^{\infty}$ curve $\gamma$ in $\U$ whose image is a 
	closed topological subset of $\M$.
	$($The latter condition excludes the case that $\gamma$ has an
	end-point on $\partial \U$.$)$ Then the following is true.
\\
	{\em (a)} If $\gamma$ does not meet the point $p$, then there is a 
	past-pointing lightlike geodesic from $p$ to $\gamma$ that 
	lies completely within $\U$ and contains no conjugate points 
	in its interior. $($The end-point may be conjugate to the 				initial-point.$)$ If this geodesic meets $\gamma$ at the point 
	$q$, say, then all points on $\gamma$ that lie to the future 
	of $q$ cannot be reached from $p$ along a past-pointing lightlike 			geodesic in $\U$.
\\
	{\em (b)} If $\gamma$ meets neither the point $p$ nor the caustic of 
	the past light cone of $p$, then the number of past-pointing 
	lightlike geodesics from $p$ to $\gamma$ that are completely 
	contained in $\U$ is finite and odd.
\end{proposition}
\begin{proof}
	In the first step we construct a $C^{\infty}$ vector field $V$ on 
	$\M$ that is timelike on $\U$, has $\gamma$ as an integral curve,
	and coincides with $W$ on $\T = \partial \U$. To that end we first 			choose any future-pointing timelike $C^{\infty}$ vector field 
	$V_1$ on $\M$. (Existence is guaranteed by our assumption of 				time-orientability.) Then we extend the vector field $W$ to a 				$C^{\infty}$ vector field $V_2$ onto some neighborhood $\V$ of 
	$\T$. Since $W$ is causal and future-pointing, $V_2$ may be 
	chosen timelike and future-pointing on $\V \setminus \T$. (Here 
	we make use of the fact that $\T = \partial \U$ is a closed 
	subset of $\M$.) Finally we choose a timelike and future-pointing 
	vector field $V_3$ on some neighborhood $\W$ of $\gamma$ that is 
	tangent to $\gamma$ at all points of $\gamma$. (Here we make use 
	of the fact that the image of $\gamma$ is a closed subset of 
	$\M$.) We choose the neighborhoods $\V$ and $\W$ disjoint which 
	is possible since $\gamma$ is completely contained in $\U$ and 
	closed in $\M$. With the help of a partition of unity we may 
	now combine the three vector fields $V_1, V_2, V_3$ 
	into a vector field $V$ with the desired properties. 

	In the second step we consider the quotient space $\M /V$. This 
	space contains the open subset $\U /V$ whose boundary $\T /V = \N$
	is, by Proposition~\ref{prop:sln}, a manifold diffeomorphic to $S^2$.
	We want to show that $\U /V$ is a manifold (which, according to our
	terminology, in particular requires that $\U /V$ is a Hausdorff space).  	To that end we consider the map $j_p : 
	{\overline{\partial _{\U} J^-(p, \U)}} \longrightarrow 
	{\overline{\U}} /V$ which assigns to each point $q \in 
	{\overline{\partial _{\U} J^-(p, \U)}}$
	the integral curve of $V$ passing through that point. 
	(Overlining always means closure in $\M$.) Clearly, $j_p$ 
	is continuous with respect 
	to the topology ${\overline{\partial _{\U} J^-(p, \U )}}$ inherits as a 
	subspace of $\M$ and the quotient topology on ${\overline{\U}} /V$. 
	Moreover, ${\overline{\partial _{\U} J^- (p, \U )}}$
	intersects each integral curve of $V$ at most once, and if it 
	intersects one integral curve then it also intersects all 
	neighbboring integral curves in ${\overline{\U}}$; this follows
	from Wald \cite{Wa}, Theorem~8.1.3. Hence, $j_p$ is injective and
	its image is open in ${\overline{\U}} /V$. On the other hand,
	part (b) of Lemma~\ref{lem:boundary} implies that the image 
	of $j_p$ is closed. Since the image of $j_p$ is non-empty and 
	connected, it must be all of ${\overline{\U}} /V$. (The domain 
	of $j_p$ and, thus, the image of $j_p$ is non-empty because 
	$\U$ does not contain a closed timelike 
	curve. The domain and, thus, the image of $j_p$ is 
	connected since $\U$ is connected.)
	We have, thus, proven that $j_p$ is a homeomorphism.
	This implies that the Hausdorff condition is satisfied on 				${\overline{\U}} /V$ and, in particular, on $\U /V$. Since 
	$V$ is timelike and $\U$ contains no closed timelike curves, 
	this makes sure that $\U /V$ is a manifold according to our 
	terminology, see Harris \cite{Ha}, Theorem~2. 

	In the third step we use these results to prove part (a) of 
	the proposition. Our result that $j_p$ is a homeomorphism 
	implies, in particular, that $\gamma$ has an intersection with 
	$\partial _{\U} J^- (p, \U)$ at some point $q$. Now part (a) of
	Lemma~\ref{lem:boundary} shows that there is a 
	past-pointing lightlike geodesic from $p$ to $q$ in $\U$. 
	This geodesic cannot contain conjugate points in its interior 
	since otherwise a small variation would give a timelike curve 
	from $p$ to $q$, see Hawking and Ellis \cite{HE}, 
	Proposition~5.4.12, thereby contradicting 
	$q \in \partial _{\U} J^-(p, \U)$. The rest 
	of part (a) is clear since all past-pointing lightlike geodesics 
	in $\U$ that start at $p$ are confined to $J^-(p, \U )$.

	In the last step we prove part (b). To that end we choose on the 
	tangent space $T_p \M$ a Lorentz basis $(E^1_p,E^2_p,E^3_p,E^4_p)$
	with $E^4_p$ future-pointing, and we identify each 
	$x = (x^1,x^2,x^3) \in \R^3$ with the past-pointing lightlike 
	vector $Y_p = x^1E^1_p + x^2E^2_p + x^3E^3_p - |x| E^4_p$. 
	With this identification, the lens map takes the form $f_p :
	S^2 \longrightarrow \N =  \partial \U /V \, , \; 
	x \longmapsto \pi _V \big( {\mathrm{exp}}_p ( w_p (x) x ) \big)$. 
	We now define a continuous map $F: B \longrightarrow \M /V$ on 
	the closed ball $B = \big\{ x \in \R ^3 \, \big| 
	\, |x| \le 1 \, \big\}$ by setting $F(x) = 
	\pi _V \big( {\mathrm{exp}}_p ( w_p (\tfrac{x}{|x|}) \, x ) \big)$
	for $x \neq 0$ and $F(0) = \pi _V (p)$.  
	The restriction of $F$ to the interior of $\B$ is a $C^{\infty}$ 
	map onto the manifold $\U /V$, with the exception of the origin 
	where $F$ is not differentiable. The latter problem can be 
	circumvented by approximating $F$ in the $C^o$-sense, on an 
	arbitrarily small neighborhood of the origin, by a $C^{\infty}$ 
	map. Then the mapping degree $\g (F)$ can be calculated (see, 
	e.g., Choquet-Bruhat, Dewitt-Morette, and Dillard-Bleick 
	\cite{CDD}, pp.477) with the help of the integral formula
\begin{equation}\label{eq:omega}
	\int _B F^* \omega = \g(F) \, \int_{\U /V} \omega
\end{equation}
	where $\omega$ is any 3-form on $\U /V$ and the star denotes the
	pull-back of forms. For any 2-form $\psi$ on $\U /V$, we may
	apply this formula to the form $\omega = d \psi$. With the help
	of the Stokes theorem we then find 
\begin{equation}\label{eq:Stokes}
	\int _{S^2} F^* \psi = \g(F) \, \int_{\N} \psi \, .
\end{equation}
	However, the restriction of $F$ to $\partial B = S^2$ gives the
	lens map, so on the left-hand side of (\ref{eq:Stokes}) we may 
	replace $F^* \psi$ by $f_p^* \psi$. Then comparison with the 
	integral formula for the degree of $f_p$ shows that $\g (F) = 
	\g (f_p)$ which, according to Proposition~\ref{prop:sln}, is 
	equal to $\pm 1$. For every $\zeta \in \U /V$ that is a regular 
	value of $F$, the result $\g(F) = \pm 1$ implies that the number 
	of elements in $F^{-1} (\zeta )$ is finite and odd. By assumption, 
	the worldline $\gamma \in \U /V$ meets neither the point $p$
	nor the caustic of the past light cone of $p$. The first 
	condition makes sure that our perturbation of $F$ near the 
	origin can be done without influencing the set $F^{-1} (\gamma )$; 
	the second condition implies that $\gamma$ is a regular value of
	$F$, please recall our discussion at the end of 
	Section~\ref{sec:regular}. This completes the proof.
\end{proof}
 
	If only light rays within $\U$ are taken into account,
	then Proposition~\ref{prop:gamma} can be summarized by
	saying that, for light sources in a simple lensing
	neighborhood, the ``youngest image'' has always even parity
	and the total number of images is finite and odd.

	In the quasi-Newtonian approximation formalism it is a
	standard result that a transparent gravitational lens
	produces an odd number of images, see Schneider, 
	Ehlers and Falco \cite{SEF}, Section~5.4, for a detailed
	discussion. Proposition~\ref{prop:gamma} may be viewed as a
	reformulation of this result in a Lorentzian geometry setting.
	It is quite likely that an alternative proof of 
	Proposition~\ref{prop:gamma} can be given by using the Morse theoretical results 
	of Giannoni, Masiello and Piccione \cite{GMP1,GMP2}. Also,
	the reader should compare our results with the work of McKenzie
	\cite{McK} who used Morse theory for proving an odd-number
	theorem in certain globally hyperbolic spacetimes. Contrary to 
	McKenzie's theorem, our Proposition~\ref{prop:gamma} requires
	mathematical assumptions which can be physically interpreted
	rather easily.  

%%%%%%%%%%%%%%%%%%%%%%%%%%%%%%%%%%%%%%%%%%%%%%%%%%%%%%%%%%%%%%%%%%%%%%%%%%%

\section{Examples}\label{sec:examples}
\subsection{Two simple examples with non-transparent 
deflectors}\label{subsec:string}

\noindent
	{\bf (a) Non-transparent string}

	As a simple example, we consider gravitational lensing in the
	spacetime $(\M,g)$ where $\M = \R^2 
        \times \big( \R^2 \setminus \{0\} \big)$ and
\begin{equation}\label{eq:string}
        g = -dt^2 + dz^2 + dr^2 + k^2\, r^2 \, d{\varphi}^2 
\end{equation}
        with some constant $0 < k < 1$. Here $(t,z)$ 
        denote Cartesian coordinates on $\R^2$ and $(r,\varphi )$
        denote polar coordinates on $\R^2 \setminus \{0\}$.
        This can be interpreted as the spacetime around a static
        non-transparent string, see Vilenkin \cite{Vi}, Hiscock \cite{Hc} 
        and Gott \cite{Go}. 
        One should think of the string as being situated at the $z$-axis. 
        Since the latter is not part of the spacetime, it is indeed 
        justified to speak of a {\em non-transparent\/} string. 

	As $\partial / \partial t$ is a Killing vector field normalized
	to $-1$, the lightlike geodesics in $(\M,g)$ correspond to the
	geodesics of the space part. The latter is a metrical product
	of a real line with coordinate $z$ and a cone with polar 
	coordinates $(r,\varphi)$. So the geodesics are straight lines
	if we cut the cone open along some radius $\varphi = {\mathrm{const.}}$
	and flatten it out in a plane. Owing to this simple form of the
	lightlike geodesics, the investigation of lens maps in this 
	string spacetime is quite easy.   

	To work this out, choose some constant $R>0$ and let 
	$\T$ denote the hypercylinder $r = R$ in $\M$. Let $W$ denote 
	the restriction of the vector field $\partial / \partial t$ to 
	$\T$. Then $(\T,W)$ is a source surface, with $\N = \T /W \simeq 
	S^1 \times \R$. Henceforth we discuss the lens map $f_p$ for any
	point $p \in \M$ at a radius $r < R$. There are no past-pointing
	lightlike geodesics from $p$ that intersect $\T$ more than 
	once or touch $\T$ tangentially, so the lens map $f_p$ gives 
	full information about all images at $p$ of each light source 
	$\xi \in \N$. The domain $\D_p$ of the lens map is given by 
	excising a curve segment, namely a meridian including both 
	end-points at the ``poles'', from the celestial sphere $\s_p$, 
	so $\D_p \simeq \R^2$ is connected. The boundary of $\D_p$ in 
	$\s_p$ corresponds to light rays that are blocked by the 
	string before reaching $\T$. It is easy to see that the 
	lens map cannot be continuously extended onto $\s_p$ (=
	closure of $\D_p$ in $\s_p$). Nonetheless, the lens map admits
	an extension in the sense of Definition~\ref{def:extend}. We
	may choose $\M_1 = S^2$ and $\M_2 = S^2$. Here $\D_p$ is embedded
	into the sphere in such a way that it covers a region
	$(\theta , \varphi ) \in \; ]0,\pi [ \; \times \;
	]\, \varepsilon \, ,  2 \pi - \varepsilon [ \,$, 
	i.e., in comparison with the embedding into $\s_p$ 
	the curve segment excised from the sphere has been
	``widened'' a bit. The embedding of $\N \simeq S^1 \times \R$ into
	$S^2$ is made via Mercator projection.

	As the string spacetime has vanishing curvature, the light cones
	in $\M$ have no caustics. Owing to our general results
	of Section~\ref{sec:regular}, this implies that the caustic 
	of the lens map is empty and that all images have even parity, 
	so (\ref{eq:degn}) gives $\g ( {\overline{f_p}} , \xi ) = 
	n_+ (\xi) = n(\xi)$ for all $\xi \in \N \setminus 
	{\overline{f_p}}(\partial \D_p )$.

	The actual value of $n (\xi)$ depends on the parameter $k$ that 
	enters into the metric (\ref{eq:string}). If $i = 1/k$ is an
	integer, $\N \setminus {\overline{f_p}}(\partial \D_p)$ is
	connected and $n (\xi) = i$ everywhere on this set. If $i <
	1/k < i+1$ for some integer $i$, $\N \setminus 
	{\overline{f_p}}(\partial \D_p)$ has two connected components,
	with $n (\xi ) = i$ on one of them and $n (\xi ) = i+1$
	on the other. Thus, the string produces multiple imaging and
	the number of images is (finite but) arbitrarily large if
	$k$ is sufficiently small. 

	For all $k \in \; ]0,1[ \,$, the lens map is surjective, 
	$f_p (\D_p) = \N \simeq S^1 \times \R$. So this example 
	shows that the assumption of $f_p(\D_p)$ being simply connected 
	was essential in Proposition~\ref{prop:cover}. 

\vspace{0.2cm}
\noindent
	{\bf (b) Non-transparent spherical body}

	We consider the Schwarzschild metric
\begin{equation}\label{eq:ss}
	g = \big( \, 1 \, - \tfrac{2m}{r} \, \big)^{-1} dr^2 + 
	r^2 \big( d \theta ^2 + 
	{\mathrm{sin}}^2 \theta \, d \varphi ^2 \big) - 
	\big( \, 1 \,  - \tfrac{2m}{r} \, ) \, dt^2 
\end{equation}
	on the manifold $\M = \; ]R_o,\infty [ \; \times \, S^2 
	\times \R$. In (\ref{eq:ss}), $r$ is the coordinate
	ranging over $\, ]R_o, \infty [ \;$, $t$ is the 
	coordinate ranging over $\R$, and $\theta$ and $\varphi$
	are spherical coordinates on $S^2$. This gives the static 
	vacuum spacetime around a spherically symmetric body of 
	mass $m$ and radius $R_o$. Restricting the spacetime
	manifold to the region $r > R_o$ is a way of treating the 
	central body as non-transparent. In the following we keep a 
	value $R_o > 0$ fixed and we allow $m$ to vary between 
	$m = 0 $ (flat space) and $m = R_o/2$ (black hole). 

	For discussing lens maps in this spacetime we fix a constant
	$R > 3 R_o /2$. We denote by $\T$ the set of all points
	in $\M$ with coordinate $r = R$ and we denote by $W$ the 
	restriction of $\partial / \partial t$ to $W$. Then 
	$(\T ,W)$ is a source surface, with $\N = \T /W \simeq S^2$. 
	It is our goal to discuss the properties of the lens
	map $f_p : \D_p \longrightarrow \N$ for a point $p \in \M$ 
	with a radius coordinate $r < R$ in dependence of the
	mass parameter $m$. To that end we make use of well-known
	properties of the lightlike geodesics in the Schwarzschild 
	metric, see, e.g., Chandrasekhar \cite{Ch}, Section~20, for 
	a comprehensive discussion. For determining the relevant
	features of the lens map it will be sufficient to concentrate 
	on qualitative aspects of image positions. For quantitative 
	aspects the reader may consult Virbhadra and Ellis \cite{VE}.
 
	We first observe that, for any $m \in [0,R_o/2]$, there is 
	no past-pointing lightlike geodesic from $p$ that intersects 
	$\T$ more than once or touches $\T$ tangentially. This 
	follows from the fact that in the region $r > 3m$ the 
	radius coordinate has no local maximum along any light ray. 
	So the lens map $f_p$ gives full information about all 
	images at $p$ of light sources $\xi \in \N$. 

\begin{figure}
\begin{center}
\setlength{\unitlength}{1cm}
\begin{picture}(10,10)
	\put(5,5){\bigcircle{8}}
	\put(5,5){\bigcircle{3.7}}
	\put(5,9){\circle*{0.1}}		
	\put(5,1){\circle*{0.1}}
	\put(5,8){\circle*{0.1}}
	\curve(5,8, 3,4.4, 5,1)
	\curve(5,8, 7,4.4, 5,1)
	\put(5.4,8.1){$p$}
        \put(4.8,9.3){$\xi_N$}
	\put(4.8,0.5){$\xi_S$}
	\put(4.3,3.6){$r = R_o$}
	\put(7.5,1.6){$r = R$}
\end{picture}
\end{center}
\caption{At $m = m_1$, the extended lens map 
	${\overline{f_p}}$ maps the boundary
	of $\D_p$ onto the south pole $\xi _S$.}\label{fig:ss}
\end{figure}
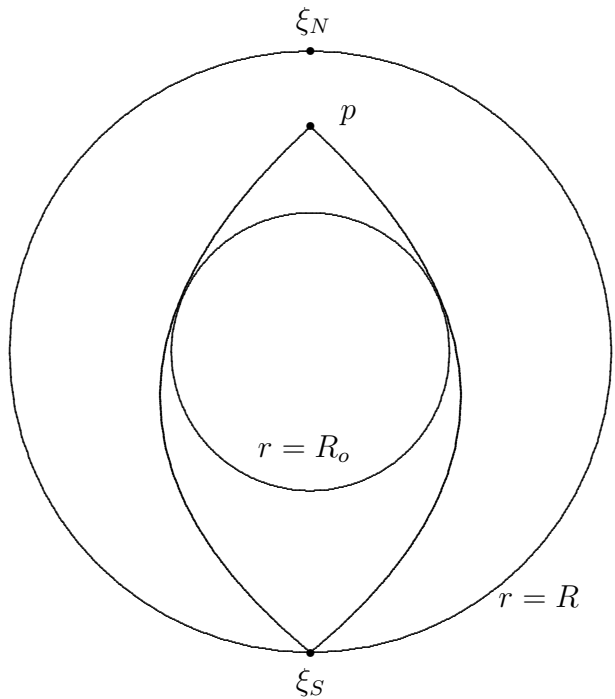

	For $m=0$, the light rays are straight lines. The domain 
	$\D_p$ of the lens map is given by excising a disc, including 
	the boundary, from the celestial sphere $\s_p$, i.e., $\D_p 
	\simeq \R^2$. The boundary of $\D_p$ corresponds
	to light rays grazing the surface of the central body, so
	$f_p$ can be continuously extended onto the closure of $\D_p$
	in $\s_p$, thereby giving an extension of $f_p$, in the sense
	of Definition~\ref{def:extend}, ${\overline{f_p}} :
	{\overline{\D_p}} \subseteq \s_p \longrightarrow \N$. 
	In Figure~\ref{fig:ss}, ${\overline{f_p}}(\partial \D_p)$ can
	be represented as a ``circle of equal latitude'' on the
	sphere $r=R$, with the image of $f_p$ ``to the north'' of 
	this circle. With increasing $m$, 
	${\overline{f_p}}(\partial \D_p)$ moves ``south'' until, 
	at some value $m = m_1$, it has reached the ``south pole'' 
	$\xi_S$. This is the situation depicted in Figure~\ref{fig:ss}.
	From now on the lens map is surjective and $\xi_S$ 
	is seen as an {\em Einstein ring}, thereby indicating that
	a caustic has formed. Now ${\overline{f_p}} (\partial \D_p)$ 
	moves north until, at some value $m = m_1'$, it has reached the 
	``north pole'' $\xi_N$. From now on $\xi_N$ is seen as an 
	Einstein ring, in addition to the regular image that exists 
	from the beginning. With further increasing $m$, we find an 
	infinite sequence of values $0 < m_1 < m_1' < m_2 < m_2' < 
	\dots < m_i < m_i' < \dots $ such that at $m = m_i$ a new 
	Einstein ring of $\xi_S$ and at $m_i'$ a new Einstein ring
	of $\xi_N$ comes into existence. For all intermediate values
	of $m$, ${\overline{f_p}}(\partial \D_p)$ divides $\N$ into two 
	connected components. All points $\xi$ in the southern component, 
	with the exception of the south pole $\xi_S$, are regular values 
	of the lens map. $f_p^{-1}(\xi )$ consists of exactly $2i$ 				points where $i$ is the largest integer with $m_i < m$. There are
	$i$ images of even parity, $n_+(\xi) = i$, and $i$ images of 
	odd parity, $n_-(\xi) = i$, hence $\g ({\overline{f_p}}, \xi )
	= n_+(\xi) - n_-(\xi) = 0$. Similarly, all points $\xi$ in the 
	northern component, with the exception of the north pole $\xi_N$, 
	are regular values of the lens map. $f_p^{-1}(\xi )$
	consists of exactly $2i+1$ points, where $i$ is the largest integer
	with $m_i' < m$. There are $i+1$ images of even parity,
	$n_+(\xi) = i+1$, and $i$ images of odd parity, $n_-(\xi) = i$,
	hence $\g ({\overline{f_p}} , \xi ) = n_+(\xi) - n_-(\xi) = 1$. Both
	sequences $(m_i)_{i \in \n}$ and $(m_i')_{i \in \n}$ converge
	towards $m = R_o/3$. For $m \ge R_o/3$, the boundary of $\D_p$ 
	corresponds to light rays that approach the sphere $r = 3m$
	asymptotically in a neverending spiral motion, cf. Chandrasekhar
	\cite{Ch}, Figure~9 and Figure~10. The lens map no  longer admits 
	an extension in the sense of Definition~\ref{def:extend},
	so we cannot assign a mapping degree to it. There are infinitely
	many concentric Einstein rings for both poles, and infinitely
	many isolated images for all other $\xi \in \N$, with both
	$n_+(\xi) $ and $n_-(\xi)$ being infinite. These features
	remain unchanged until the black-hole case $m = R_o/2$ is reached.

	The fact that in this case the caustic of the lens map consists 
	of just two points is rather exceptional. After a small perturbation 
	of the spherical symmetry the caustic would show a completely 
	different behavior. For regular $\xi \in \N$, however, the 
	statements about $n_{\pm}(\xi )$ are stable against small
	perturbations. 

	Having studied Schwarzschild spacetimes around non-transparent
	bodies, the reader might ask what about transparent bodies,
	i.e., what about matching an interior solution to the
	exterior Schwarzschild solution at Radius $R_o$, with $R_o > 2m$,
	and allowing for light rays passing through the interior
	region. If $R_o > 3m$, and if there are no light rays trapped
	within the interior region, the resulting spacetime will be 
	asymptotically simple and empty. Qualitative features of lens maps
	in this class of spacetimes are discussed in the following subsection. 
	For a more explicit discussion of lens maps in the Schwarzschild
	spacetime of a transparent body, choosing a perfect fluid with 
	constant density for the interior region, the reader is refered
	to Kling and Newman \cite{KNm}.

%-----------------------------------------------------------------------------

\subsection{Asymptotically simple and empty spacetimes}\label{subsec:asy}
	Asymptotically simple and empty spacetimes are considered to
	be good models for the gravitational fields of transparent 
	gravitating bodies that can be viewed as isolated from 
	all other masses in the universe. The formal definition, which
	is essentially due to Penrose \cite{Pn}, cf., e.g. Hawking 
	and Ellis \cite{HE}, p. 222, reads as follows. 

\begin{definition}\label{def:asy}
        A spacetime $(\M,g,)$ is called 
        {\em asymptotically simple\/} if there is a strongly causal 
        spacetime $({\tilde{\M}},{\tilde{g}})$ with the following 
	properties.
\\
	(a) $\M$ is an open submanifold of ${\tilde{\M}}$ with a 
	non-empty boundary $\partial \M\,$. 
\\
	(b) There is a $C^{\infty}$ function $\Omega : 
        {\tilde{\M}} \longrightarrow {\mathbb{R}}$ 
        such that $\M = 
        \{ \, p \in {\tilde{\M}} \,  | \; \Omega (p) > 0 \; \}$, 
        $\partial \M = 
        \{ \, p \in {\tilde{\M}} \, | \, \Omega (p) = 0 \, \}$, 
	$d \Omega \neq 0$ everywhere on $\partial \M$ and
        ${\tilde{g}} = \Omega^2 \, g \,$
        on $\M\,$.
\\
	(c) Every inextendible lightlike geodesic in ${\mathcal{M}}$ has 
        past and future end-point on $\partial \M \,$.
\\                                           
        $(\M,g)$ is called 
        {\em asymptotically simple and empty\/} if, in addition,
\\
	(d) there is a neighborhood $\V$ of $\partial \M$ in 							${\tilde{\M}}$ such that the Ricci tensor of $g$ vanishes 
	on $\V \cap \M$.
\end{definition}
      
	Condition (d) of Definition~\ref{def:asy} is a way of saying 
	that, sufficiently far away from the gravitating body under 						consideration, Einstein's vacuum field equation is satisfied. 
	This assumption is reasonable for the spacetime around an
	isolated body producing gravitational lensing as long as
	cosmological aspects can be ignored. 
         
        The assumptions (a)--(d) of Definition~\ref{def:asy}
        imply that $\partial \M$ is a
        $\tilde{g}$-lightlike hypersurface 
        in ${\tilde{\M}}$ that has two connected components, 
	usually denoted by $\J^+$ and $\J^-$ (cf., e.g., Hawking
	and Ellis, \cite{HE}, p.222). 
	Every inextendible lightlike geodesic in $\M$ has future
	end-point on  $\J^+$ and past end-point on 
	$\J^-$. 

	In the following we concentrate on $\J^-$ which
	is the relevant quantity in view of gravitational lensing.
	By construction, $\J^-$ is ruled by the integral
	curves of the ${\tilde{g}}$-gradient $Z$ of $\Omega$. (In
	coordinate notation, the vector field $Z$ is defined by
	$Z^a = {\tilde{g}}{}^{ab} \, \partial _b \Omega$ on 
	$\J^-$.) It is well known that $Z$ is regular, with
	$\J^- /Z$ being diffeomorphic to $S^2$, and that the 
	natural projection $\pi _Z : \J^- \longrightarrow 
	\J^- /Z \simeq S^2$ makes $\J^-$ into a trivializable
	fiber bundle with typical fiber diffeomorphic to $\R$.
	For a full proof we refer to Newman and Clarke \cite{NC,Ne}.
	(The argument given in Hawking and Ellis \cite{HE}, 
	Proposition~6.9.4, which is due to Geroch \cite{Ge}, is
	incomplete.) This result can be translated into our 
	terminology in the following way.
\begin{proposition}\label{prop:scri}
	In the case of an asymptotically simple and empty 
	spacetime, $(\J^-  , Z)$ is a source surface in the 
	spacetime $({\tilde{\M}}, {\tilde{g}})$, with 
	$\N = \J^- /Z$ diffeomorphic to $S^2$.
\end{proposition}
	Each integral curve of $Z$ can be written as the $C^1$-limit
	of a sequence $(\gamma _i)_{i \in \n}$ of timelike curves 
	in $\M$. We may interpret the $\gamma _i$ as a sequence
	of worldlines of light sources approaching infinity. 
	From the viewpoint of the physical spacetime $(\M,g)$,
	it is thus justified to interpret the integral curves of 
	$Z$ as ``light sources at infinity''. With respect to the
	unphysical metric ${\tilde{g}}$, these worldlines are 
	lightlike. With respect to the physical metric, however,
	they have no causal character at all, because the metric
	$g$ is not defined on $\J^-$. It is, thus, a misinterpretation
	to say that the ``light sources at infinity'' move at the
	speed of light.
	
	We shall now show that the formalism of ``simple lensing
	neighborhoods'' applies to the situation at hand. To that
	end, we observe that $\J^-$ is the boundary of $\M$ in the 
	manifold ${\tilde{M}} \setminus \J^+$. This gives rise to 
	the following result.
\begin{proposition}\label{prop:asln}
	In the case of an asymptotically simple and empty spacetime,
	$(\M, \J^-,Z)$ is a simple lensing neighborhood in the
	spacetime $({\tilde{\M}} \setminus \J^+ , {\tilde{g}}|
	_{{\tilde{\M}} \setminus \J^+})$.
\end{proposition}
\begin{proof}
	Condition (a) of Definition~\ref{def:sln} is obvious from
	Definition~\ref{def:asy} and Condition (b) was just
	established. The proof of the remaining two conditions 
	is based on the fact that on $\M$ the $g$-lightlike geodesics 
	coincide with the ${\tilde{g}}$-lightlike geodesics (up to 
	affine parametrization). Condition (d) of 
	Definition~\ref{def:sln} is satisfied since every lightlike 
	geodesic in $\M$ has past end-point on $\J^-$ and
	future end-point on $\J^+$. Moreover, the arrival on
	$\J^{\pm}$ must be transverse since $\J^{\pm}$ is
	${\tilde{g}}$-lightlike. This shows that Condition
	(c) of Definition~\ref{def:sln} is satisfied as well.
\end{proof}

	We can, thus, apply our results on simple lensing neighborhoods
	to asymptotically simple and empty spacetimes. As a first
	result, Proposition~\ref{prop:sln} tells us that every
	asymptotically simple and empty spacetime $\M$ must be 
	contractible. This result is not new. It
	is well known that every asymptotically simple and empty
	spacetime is globally hyperbolic and, thus, homeomorphic
	to a product of a Cauchy surface $\C$ with the real line,
	$\M \simeq \C \times \R$, and that $\C$ is contractible.
	For a full proof we refer again to Newman and Clarke
	\cite{NC,Ne}. The stronger result that $\C$ must be
	homeomorphic to $\R^3$ requires the assumption that the
	Poincar\' e conjecture is true (i.e., that every simply 
	connected and compact 3-manifold is homeomorphic to $S^3$).

	In addition, Proposition~\ref{prop:sln} gives us the 
	following result.
\begin{proposition}\label{prop:asyb}
	In the case of an asymptotically simple and empty spacetime,
	for all $p \in \M$ the lens map $f_p :
	\s_p \longrightarrow \J^- /Z \simeq S^2$ has $| \g (f_p) |= 1$.
\end{proposition}
	
	The lens map $f_p$ for ``light sources at infinity''
	in an asymptotically simple and empty spacetime was 
	already discussed in Perlick \cite{Pe1,Pe2}. In particular,
	a proof of the result $\g (f_p) = 1$ was given in 
	Theorem~6 of \cite{Pe1}. An equivalent statement, using
	a different terminology, can be found as Lemma~1 in 
	Kozameh, Lamberti and Reula \cite{KLR}, together with 
	a short proof. However, both these earlier proofs are 
	incomplete. The proof in \cite{Pe1} is based on the idea 
	to homotopically deform $f_p$ into the identity, but it 
	is not shown that the construction can be made in such a 
	way that the dependence on the deformation parameter is, 
	indeed, continuous. In \cite{KLR}, the authors write the future 
	light cone (or, equivalently, the past light cone) of a 
	point $p \in \M$ as the image of a map $\Phi : \: ]0, \infty[
	\: \, \times S^2 \longrightarrow \M$, and they assign a 
	{\em winding number\/} to each map $\Phi (s, \cdot )$. 
	Since a winding number has to refer to a ``center'', the
	authors in \cite{KLR} apparently take for granted that 
	there is a timelike curve through $p$ that has no further 
	intersection with the light cone of $p$. The existence of 
	such a curve, however, is an open question. With our 
	Proposition~\ref{prop:sln} we have filled these gaps insofar as 
	we have established the result $\g (f_p) = \pm 1$. However, we 
	have not shown whether, with our choice of orientations,
	the occurence of the minus sign can be ruled out.
        
\begin{figure}
\begin{center}
\setlength{\unitlength}{0.8cm}
\begin{picture}(8,9)
	\curve(2,1, 2,9)
	\curve(2,1, 6,5)
	\curve(6,5, 2,9)
	\curve(2,1, 3.5,4, 4,7)
	\put(3,6.5){\circle*{0.08}}
	\put(2.5,6.5){$p$}
        \put(4,4.5){$\gamma$}
	\put(4.3,2.5){$\J^-$}
	\put(4.3,7.5){$\J^+$}
\end{picture}
\end{center}
\caption{Illustration of Proposition~\ref{prop:asygamma}}\label{fig:asy} 
\end{figure}
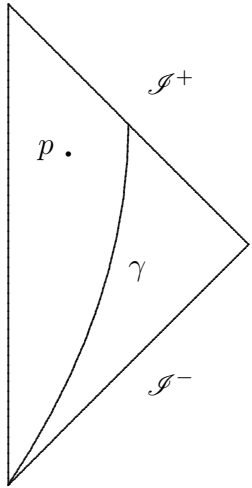

	Proposition~\ref{prop:asyb} implies that every observer in 
	$p$ sees an odd number of images of each light source at
	infinity that does not pass through the caustic of the 
	past light cone of $p$. (Here one has to refer to the 
	${\tilde{g}}$-cone which is an extension of the $g$-cone.)
	As an immediate consequence of Proposition~\ref{prop:gamma},
	we find that a similar statement is true for light sources
	inside $\M$, see Figure~\ref{fig:asy}.
\begin{proposition}\label{prop:asygamma}
	Fix a point $p$ and a timelike embedded $C^{\infty}$curve 
	$\gamma$ in an asymptotically simple and empty spacetime $(\M,g)$. 
	Assume that the image of $\gamma$ is a closed subset of 
	${\tilde{\M}} \setminus \J ^+$ and that $\gamma$ meets neither 
	the point $p$ nor the caustic of the past light cone of $p$. 
	Then the number of past-pointing lightlike geodesics from $p$ 
	to $\gamma$ in $\M$ is finite and odd.
\end{proposition}

	Let us conclude this subsection with a few remarks on 
	spacetimes that are asymptotically simple but not empty.
	For any asymptotically simple spacetime it is easy to 
	verify that $\partial \M$ has either one or two connected
	components, and that all lightlike geodesics in $\M$ have 
	their past end-point in the same connected component of 
	$\partial \M$. Let us denote this component by $\J^-$ 
	henceforth. In order to apply our formalism of simple
	lensing neighborhoods the additional assumptions needed 
	are that $\J^-$ is a fiber bundle with ${\tilde{g}}$-causal 
	fibers diffeomorphic to $\R$ over an orientable basis manifold,
	and that all past-inextendible lightlike geodesics in $\M$
	meet $\J^-$ transversely. If these assumptions are satisfied,
	our results on simple lensing neighborhoods apply.
	In particular, $\J^-$ must be diffeomorphic to $S^2 \times \R$
	and $\M$ must be contractible. 

	As an interesting special case, we might modify Condition (d) 
	of Definition~\ref{def:asy} by requiring the Ricci tensor of
	$g$ to be equal to $\Lambda \, g$ near $\partial \M$ with a 
	positive or negative cosmological constant $\Lambda$. The 
	resulting spacetimes are called {\em asymptotically deSitter\/} 
	for $\Lambda > 0$ and {\em asymptotically anti-deSitter\/} for 
	$\Lambda < 0$. It was verified already by Penrose \cite{Pn} that 
	then $\partial \M$ is ${\tilde{g}}$-spacelike for $\Lambda > 0$ 
	and ${\tilde{g}}$-timelike for $\Lambda < 0$. Thus, the formalism
	of simple lensing neighborhoods is inappropriate for
	investigating asymptotically deSitter spacetimes, 
	but it may be used for the investigation of asymptotically 
	anti-deSitter spacetimes.

%-------------------------------------------------------------------------
\subsection{Weakly perturbed Robertson-Walker spacetimes}\label{subsec:RW}
	It is a characteristic feature of the lens map, as defined in this
	paper, that it is constructed by following each past-pointing 
	lightlike geodesic up to its first intersection with the source
	surface only. Further intersections are ignored, i.e., some 
	images are willfully excluded from the gravitational lensing
	discussion. In the preceding examples no such further intersections
	occured. We shall now discuss an example where they do occur
	but where it is physically well motivated to disregard them.

	To that end we start out with a spacetime $(\M,g)$ with 
	$\M = S^3 \times \R$ and 
\begin{equation}\label{eq:RW}
	g = R(t)^2 \big(-dt^2 +  d \chi ^2 + {\mathrm{sin}}^2 \chi
	( d \theta ^2 + {\mathrm{sin}} ^2 \theta \, d \phi ^2 ) \big) \, .
\end{equation}
	Here $\chi \in [0, \pi ]$, $\theta \in [0, \pi ]$ and 
	$\phi \in [0,2\pi]$ denote standard coordinates on $S^3$ (with
	the usual coordinate singularities), $t$ denotes the projection
	from $\M = S^3 \times \R$ onto $\R$, and $R : \R \longrightarrow \R$ 
	is a strictly positive but otherwise arbitrary $C^{\infty}$
	function. This is the general form of a Robertson-Walker spacetime
	with positive spatial curvature and natural topology which has
	no particle horizons. (Particle horizons are excluded by the
	assumption that the ``conformal time'' $t$ runs over all of $\R$.)
	
	Now fix a coordinate value $\chi_o \in \; ]\, 0 \, , \pi /2[ \;$ 
	and let $\U$ denote the set of all points in $\M$ whose 
	$\chi$-coordinate is smaller than $\chi_o$. Let $W$ denote
	the restriction of the vector field $\partial / \partial t$ to 
	the boundary $\partial U$. Then $(\U , \partial \U, W)$
	is a simple lensing
	neighborhood. This is easily verified using the fact that the
	lightlike geodesics in $\M$ project to the geodesics of the
	standard metric on $S^3$. Our assumptions that $t$ ranges over
	all of $\R$ and that $\chi_o < \pi /2$ are essential to make 
	sure that, for all $p \in \U$, the lens map is defined on all 
	of $\s_p$. In the case at hand, the lens map $f_p : \s_p 
	\longrightarrow \partial \U /W$ is a global diffeomorphism for 
	all points $p \in \U \,$. Actually, there are infinitely 
	many past-pointing lightlike geodesics from any fixed $p \in \U$
	to any fixed $\xi \in \partial \U /W$, but only one of them 
	reaches $\xi$ without having left $\U$. All the other ones 
	make at least a half circle around the whole universe, so 
	they will give rise to rather faint images as a consequence of
	absorption in the intergalactic medium. It is, thus, 
	reasonable to assume that only the one image which enters 
	into the lens map is actually visible. In this sense, 
	disregarding all the other light rays is physically well motivated.
	Please note that all the infinitely many images of $\xi$ are
	situated at just two points of the celestial sphere at $p \,$;
	the two brightest images cover all the other ones. 

	Now this example is boring in view of gravitational lensing
	because the lens map is a global diffeomorphism. However, we
	can switch to a more interesting situation by choosing a compact
	subset $\K \in S^3$ and modifying the metric on the set
	$\K \times \R$. In view of Einstein's field equation, this can 
	be interpreted as introducing local mass concentrations that
	act as gravitational lens deflectors. If $\K \times
	\R$ is completely contained in $\U$, and if the modification 
	of the metric is sufficiently small to make sure that, even
	after the modification, no light rays are past- or future-trapped 
	inside $\U$, then $\U$ remains a simple lensing neighborhood. We 
	have, thus, Proposition~\ref{prop:sln} at our disposal. Under the
	(very mild) additional assumption that, even after the perturbation, 			there are no closed timelike curves in $\U$, we may also
	use Proposition~\ref{prop:gamma}. This is a line of argument 
	to the effect that, in a Robertson-Walker spacetime of the 
	kind considered here, any transparent gravitational lens 
	deflector produces an odd number of visible images. The 
	assumption that there are no particle horizons was essential 
	since otherwise the lens map would not be defined on the 
	whole celestial sphere for all $p \in \U$.  

	A similar argument applies, of course, to Robertson-Walker 
	spacetimes with non-compact spatial sections. Then we don't 
	have to care about light rays traveling around the whole 
	universe, so there are no additional images which are ignored 
	by the lens map.

%%%%%%%%%%%%%%%%%%%%%%%%%%%%%%%%%%%%%%%%%%%%%%%%%%%%%%%%%%%%%%%%%%%%%%%%%%%%

\end{document}